# Observation of intrinsic chirality of surface plasmon resonances in single nanocrystals

Jong-Won Park[1,2,*]

[1]*on leave from the Department of Chemistry and The Photonics Center, Boston University, Boston, Massachusetts 02215, United States*

[2]*Department of Mechanical Engineering, Texas Tech University, Lubbock, Texas 79409, United States*

*Corresponding author: jwonpark@bu.edu (J.-W. P.)
ORCID iD: 0000-0002-1167-3230

**ABSTRACT.** Surface plasmons in a centrosymmetric metal nanocrystal have been perceived as achiral, based on the quantum-mechanical theory of molecular optical activity. The one-electron theory is insufficient to describe optical activity in crystals, where spatial variation of polarization states along the light propagation direction is crucial in dimensions exceeding molecules. Such spatial-dispersion, which can drastically modify the optical near-field in plasmonic nanocrystals, has yet to be experimentally proved at the nanometer scale. Here, the experimental observation of natural circular dichroism of surface plasmons in spherical gold, silver, and copper nanocrystals in solution is presented. It originates from spatial dispersion (nonlocality) occurred at a dimension as small as 5 nanometers in the optical region. In particular, the electromagnetic pressure or longitudinal volume plasmons result in phase-shift and polarization rotation of the transverse electric field of incident light. Dipolar surface plasmons exhibit negative ellipticity and corresponding negative phase-shift of the electric field (i.e., optically left-handed), whereas quadrupolar surface plasmons have positive signs in both ellipticity and phase-shift (i.e., optically right-handed). Structural effects such as nanocrystal assembly, nanocrystal shape, electron spin-polarization, and plasmon-molecule interaction are accessed. Volume plasmons inside the nanocrystal also produce a positive rotational strength for the interband transition. This study implicates optical symmetry-breaking in isotropic nanocrystals by the axial polarization parallel to the wavevector, which will impact on conduction electron dynamics, hot-electron generation, and nonlinear optical response at the surface of plasmonic nanocrystals as well as optical excitation of bound d-band electrons.

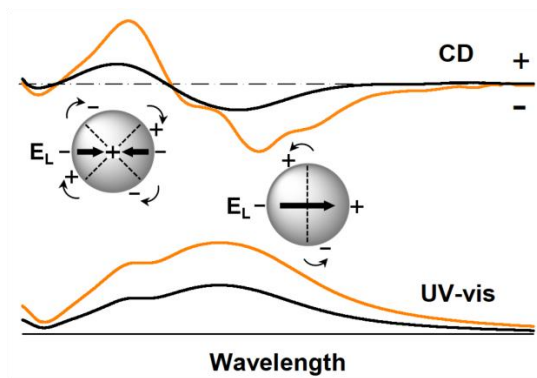

## I. INTRODUCTION

Optical activity is the consequence of helicity (chirality) in light-matter interactions [1,2,3]. The electric field vector of linearly polarized light rotates as the light propagates through a medium without having a center of symmetry. The non-centrosymmetric structure is required to generate optical activity in organic molecules and metal complexes. Such quantum-mechanical theory of molecular optical activity, based on one-electron model formulating the scalar product of electric and magnetic dipole transition moments Im{μ·m}, is well-established [1,2]. Over the past decade, this theory has been adapted in the studies of optical activity of metal nanocrystals [4,5,6], pertaining to enantiomorphism in crystal shapes and adsorbate molecules as well as helical assembly of nanocrystals. However, chiral morphology is not the requirement for optical activity in a crystal with its dimension exceeding small molecules. Under the effect of spatial dispersion, a crystal with its mirror image identical can also exhibit optical activity [7,8].

Spatial dispersion (nonlocality) [9,10], how spatial variation of a local electromagnetic field influences a macroscopic light-matter interaction, has been known to be crucial for natural optical activity in crystals [7,8,11,12]. Beyond the local approximation dealing with the transverse dielectric response (i.e., frequency dispersion), optical activity in crystals is described as to whether or not the wavevector-dependent dielectric response has a center of symmetry instead of the direct relation to crystal enantiomorphism. The nonlocal dielectric theory generally includes a propagating longitudinal wave [7,8,9,13] that may break mirror symmetry and produce optical non-centrosymmetry [14]. In fact, spatial dispersion was the theoretical foundation of optical rotatory dispersion and circular dichroism in the 1920s-1930s, and later the quantum-mechanical theory of one-electron transitions in a molecule [15] reformulated the classical polarization theory of optical activity associated with helical charge displacements in matter [1,16,17]. Although nonlocal optical response [14,18,19,20] has recently gained great attention in metallic nanostructures [21,22,23,24], there has been no report on nonlocality-driven optical activity of isotropic nanoparticles either theoretically or experimentally, probably because of the prevalent misconception regarding optical activity in the solid state or the weakness of nonlocal response of nanocrystals in far-field spectroscopic measurements.

Herein, we report the first experimental results of circular dichroism (CD) in dipolar and quadrupolar surface plasmon resonances as well as interband transitions in individual gold, silver, and copper nanocrystals in a size range of 5-150 nm that do not posses any enantiomorphism in crystal shape and ligand structure. This indicates the existence of longitudinal waves, which are strong enough to vary the phase of surface plasmons in the nanocrystals. For a metallic nanocrystal, the longitudinal wave in the ultraviolet and visible spectral regions corresponds to volume plasmons [10,20,22,24, 25 , 26 , 27 , 28 ] that are known to be excited by light in electrodynamically inhomogeneous media [29,30]. The field inhomogeneity is ascribed by spatial variation in the energy transfer [31,32 ] in the individual nanocrystals [10,33,34]. Optical handedness of surface plasmons (SPs) is determined by considering a near-field coupling of volume plasmons to SPs [35,36] and consequent phase-shift of SPs. Qualitative experimental observations include a size- or plasmon-mode-dependent CD ellipticity, a sum rule for rotational strength of SPs, enhanced CD response of quadrupolar SPs via Fano resonances [37], as well as inhomogeneous internal fields leading to a rotational strength for the optical excitation of d-electrons.

## II. CD OF SPHERICAL GOLD NANOPARTICLES

Noble metal nanoparticles absorb and scatter light in visible wavelength ranges because of localized SPs. The SP resonance is an electronic excitation by the transverse electric field of incident light, making conduction electrons oscillate coherently with the applied field. When the nanoparticle size increases to about a tenth of the wavelength of light and then all of the conduction electrons cannot oscillate coherently, quadrupolar SPs are excited at a higher energy with about one-half of the conduction electrons oscillating in the opposite direction. According to the classical Mie theory considering only tangential electromagnetic fields for far-field intensity and ignoring radial field components normal to the surface [10], the electric field of SPs is ordinarily perpendicular to the magnetic counterpart, and thus no optical activity of spherical nanoparticles is expected [38,39].

Surprisingly, noticeable CD response was observed from spherical gold nanoparticles well-dispersed in aqueous solution. Figure 1 shows electronic CD and UV-vis extinction spectra of Au spheres of representative diameters in the range of 5-150 nm. The Au spheres are stabilized by citrate anions, which are achiral. Nanoparticle concentration in the solution was adjusted to absorbance of around 1.5 and 3.0 at the wavelength of plasmon resonance for each nanoparticle size, and the path-length of light was 1.0 cm (see Methods). Obtaining CD spectra at the high concentration may verify whether or not any weaker constituent of chirality exists in addition to a dominant one, and the long path-length warrants measurements of the far-field CD intensity of SPs perturbed by near-fields and internal fields. First of all, Au spheres with diameters up to 55 nm exhibit only weakly negative ellipticities in the plasmon wavelength range of 517-534 nm, nearly proportional to concentration (Figures 1a and 1b). This yields a size-dependent molar ellipticity in the range of about $-5 \times (10^8\text{-}10^{11})$ mdeg·M$^{-1}$·m$^{-1}$ that are best fit to a cubic curve [40]. The ellipticity values are at least three order of magnitude larger than in common chiral biomolecules, due to the large absorption coefficients of the nanoparticles. Only dipolar SPs are excited in these sizes of Au spheres that are mainly absorptive rather than scattering [41]. Consequently, dipolar SPs absorb right-circularly polarized light (RCP) more than left-circularly polarized light (LCP); i.e., differential absorption in CD is $\Delta\varepsilon = \varepsilon_L - \varepsilon_R$ and thus $\varepsilon_L < \varepsilon_R$. The cubic dependence of molar ellipticity on diameter indicates the CD response has the same volume-character as the dipolar SP, and thus dipolar SPs are likely perturbed by another field component inducing a negative rotational strength. The CD response is highlighted by 5-nm Au spheres (Figure 1a), which the size is too small for a multipolar SP to be excited, revealing that asymmetry is an intrinsic property of dipolar SPs, not a consequence of dipole-multipolar interference.

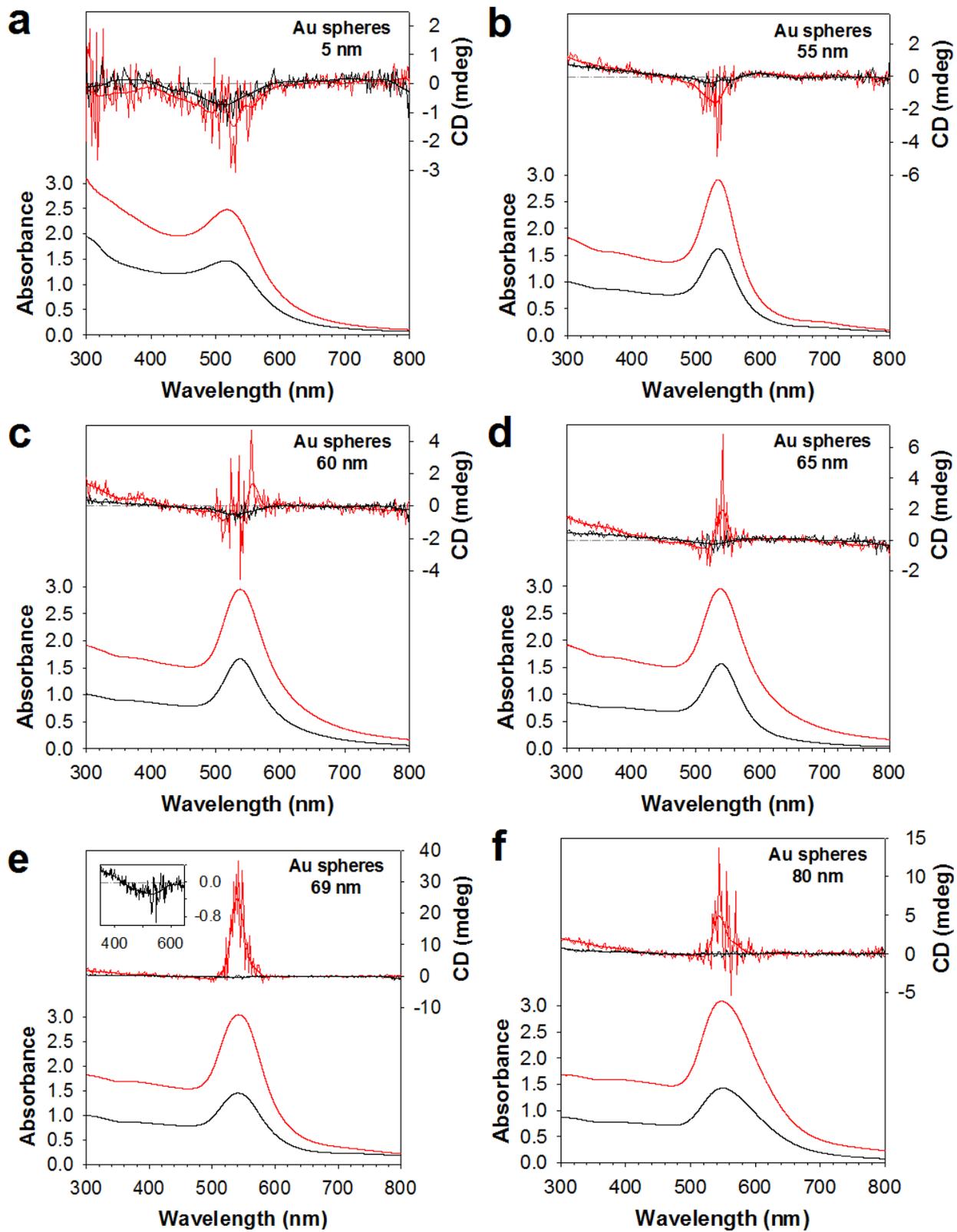

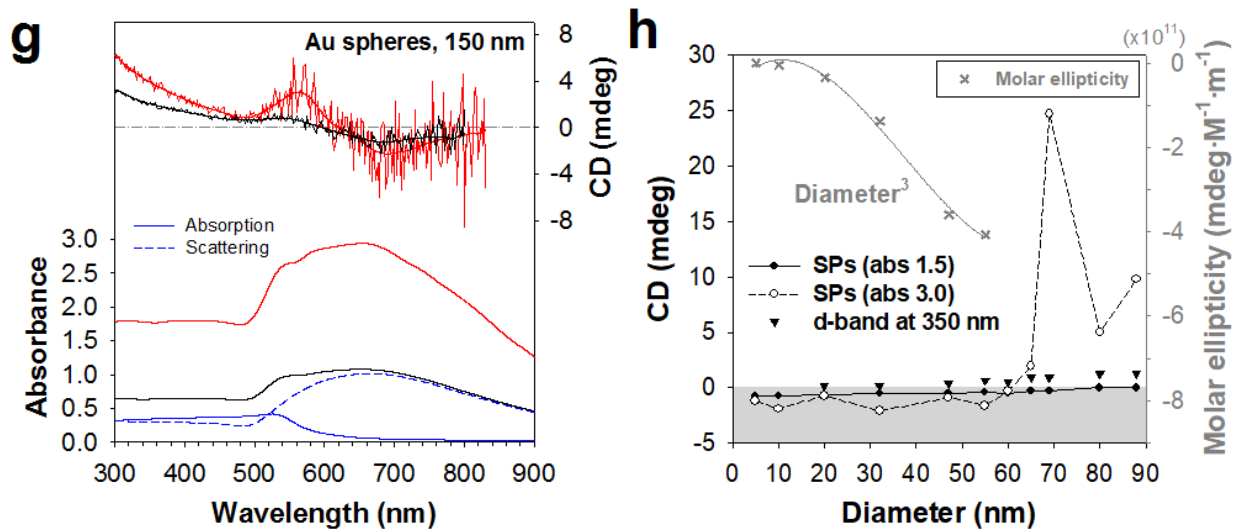

Figure 1. CD and UV-vis extinction spectra of Au spheres with different diameters. (a) 5 nm, (b) 55 nm, (c) 60 nm, (d) 65 nm, (e) 69 nm; inset: CD spectrum near the plasmon wavelength for the low concentration sample, (f) 80 nm, (g) 150 nm. Both raw and smoothened CD spectra are presented. Spectra of other sizes (10, 20, 32, 47, and 88 nm) and a magnified plot in the region of negative ellipticities are presented in the Supplemental Material. (h) summary plot of CD sign, apparent amplitude, and molar ellipticity. Path length: 1 cm.

Contrary to the small Au spheres with ellipticity sign independent of concentration, large Au spheres with a diameter range 60-88 nm change their sign from negative to positive in the plasmon wavelength range of 538-551 nm with concentration increased. Only negative ellipticities are obtained when absorbance is about 1.5, which CD of the dominating dipolar SPs is barely measurable. When absorbance increases to 3.0, for example, 60-65 nm Au spheres begin exposing a band of positive ellipticity (Figures 1c and 1d). The sign reversal indicates two types of chiral plasmons are present in the large nanoparticles, which are opposite in sign, revealing quadrupolar SPs have a positive rotational strength and absorb LCP more than RCP. This plasmon-mode-dependent sign of CD ellipticity excludes any effect of anisotropically assembled nanoparticles whose CD signals are origin-dependent and must sum to zero in solution unless one handedness in assembled nanoparticles is dominant. The CD band-width is approximately half that in UV-vis extinction spectra. Obviously, quadrupolar SPs have a lower value of molar ellipticity than the corresponding dipolar SPs, but not significantly small.

The relatively large magnitude of the positive ellipticity for the Au spheres larger than 65 nm is likely a consequence of the spectral overlap of the two different SPs. In other words, dipolar SPs having the negative rotational strength produce more fractions of LCP as the incident light passes through the sample, because the dipolar SPs not only absorb RCP but also emit LCP, which the contribution from the latter may become significant at the high concentration through multiple scattering: differential scattering $\Delta I = I_R - I_L$ [2,42]. This light-scattering effect produces an excess of LCP, i.e., LCP > 50% rather than the exact 50% generated in the CD spectrometer, resulting in the amplitude modulation of the positive ellipticity of quadrupolar SPs [40].

When the concentration of the nanoparticles is high enough to probe the CD response of quadrupolar SPs, the amplitude of the positive ellipticity of quadrupolar SPs changes non-monotonically as the nanoparticle size increases. The 65-nm Au spheres show a transition from the negative ellipticity of dipolar SPs to the positive ellipticity of quadrupolar SPs (Figure 1d). The CD band of 69-nm Au spheres marks the progress of the contribution of quadrupolar SPs with displaying an apparent ellipticity of about +25 mdeg (Figure 1e). However, the amplitude of the positive ellipticity decreases drastically as nanoparticle size increases even larger to 80 nm (Figure 1f). This nonlinear phenomenon, not observed at the low concentrations where the CD amplitude from dipolar SPs monotonically changes, may indicate the dipole-quadrupole Fano resonance [37]. Spatial overlap between dipolar and quadrupolar SPs confined in a nanoparticle is maximized at around 69 nm in the sense that overlap density of plasmon modes is the largest.

The constructive scattering-interference amplifies the CD ellipticity of quadrupolar SPs, similar to the multiple-scattering effect described above. The interference is less pronounced as nanoparticle size increases to above 80 nm.

We also obtained the first example of a sum rule for the rotational strength of conduction-electron excitations. For a given enantiomeric molecule, in general, the rotational strength or the oscillator strength sums to zero over all electronic transitions [1,2]. The 150-nm Au spheres, where conduction-electrons are excited up to a quadrupole mode and higher-order modes are negligible, display a positive CD band of absorptive quadrupolar SPs centered at a shorter wavelength of 565 nm but a negative CD band of dipolar SPs centered at a longer wavelength of 680 nm (Figure 1g). Both amplitudes yield similar molar ellipticities of the order of $10^{13}$ mdeg·$M^{-1}$·$m^{-1}$. Two isosbestic points are present in the CD spectra: one at ~490 nm, and the other at ~650 nm on the high energy side of the dipolar SP. The latter isosbestic point, because of the presence of the two SP modes, is located in the negative ellipticity region, which may result from the higher magnitude of molar ellipticity of dipolar SPs compared to quadrupolar SPs. The opposite sign of ellipticity reflects that the direction of the quadrupolar electric-field is opposite to that of the dipolar electric-field. The rotational strengths of SPs may be able to be experimentally measured by the area of the corresponding CD band [1,2]. On the basis of the CD spectrum of the Au nanoparticle sample in the absorbance 3.0, the area ratio of near unity confirms the existence of the sum rule for the oscillator strength of SPs (see Methods).

### III. VOLUME-PLASMON-INDUCED CD MECHANISM

In order to obtain accurate optical response of metal nanoparticles with considering nonlocality, one needs to add a radial-field component to the classical dielectric response [10], since the continuity of the normal field component at metal surfaces is a real physical phenomenon [26,27,35]. These radial components outward from the entire surface of a nanoparticle are generally considered as electron-density spill-out or surface dipole moment, and they may not have an effect on CD because polarity signs at both interfaces are opposite to each other regardless of whether or not the surface dipoles resonate in the visible wavelength region. Instead, the longitudinal field in a plasmonic nanoparticle is related to an excitation of longitudinal volume plasmons [20]. Volume plasmons are spherical standing waves associated with charge density fluctuations [25,28,29,30,36] or a bulk current density [43,44] inside a nanoparticle, but they extend to the surface. A transverse electric field of light cannot excite volume plasmons of free electrons in metal due to lack of a local elastic restoring force, whereas this force is generally exerted on the ions in a polar crystal [26], and thus the longitudinal volume plasmons in metal nanoparticles should be excited in the direction of light propagation.

The longitudinal volume plasmon is an indication of inhomogeneous internal field in the plasmon excitation process. Such field inhomogeneity results from spatial variation in the momentum and energy transfer from plasmons to electrons upon light absorption [10], which is also known to bring about the photon drag effect in plasmonic nanostructures [31,32]. It is most likely that the electromagnetic pressure or optical torque exerts an electromotive force [31,32] on electrons inside the nanoparticle [33,34,45,46]. Calculations by expanded Mie theory reveal the presence of optical vortex on an absorptive metal nanoparticle and near-field components in the longitudinal direction [33,34], which suggests that the axial polarizability [47] is anisotropic. The anisotropic polarizability tensor should be responsible for the optical activity of metal nanoparticles although optical vortex accompanied with a longitudinal field component [48,49] is not usually linked to optical chirality of a nanoparticle [46,50]. Longitudinal electric fields [51,52] are known to play a crucial role in the optical anisotropy of scattered electric fields observed for achiral nanostructures under normal incidence of light [51,52,53,54,55].

Since the main objective of this study is to determine handedness of SPs, we assume that coupling of longitudinal volume plasmons to SPs occur [14,20,28,29,30,36,56,57] at the nanoparticle surface in a similar excitation energy. This enables us to determine the sign of phase shift of a SP mode retrospectively by electron motion at the surface, although the realistic physical effect is that of the electromotive force whose direction may be represented by the Poynting vector. Hence, it is feasible to assume that the coupling perturbs SPs [20] and shifts the phase of the electric field of SPs [58,59,60] so that the difference of circularly polarized lights either in scattering intensity or in absorption can be measured in the far-field CD spectroscopy. On the basis of the experimental observation of the negative ellipticity of dipolar SPs that preferentially absorb RCP, the vector line of the transverse electric field in the nanoparticle *a posteriori* is shifted in the sense that the net field has a negative value of helicity (angular momentum) as is geometric phase φ- illustrated in Figure 2a. In other worlds, the dipolar SP generates a negative phase-shift of the transverse electric field, or the orbital angular momentum of dipolar SPs ($-E_z$) couples to the spin angular momentum of RCP ($\sigma_+$) [50,51]. This phase shift is different from the approximate $-\pi/2$ phase shift (time delay; out-of-phase in the imaginary absorption) of the ordinary surface plasmon resonance relative to the incident electric field [61,62,63], which the $-\pi/2$ phase shift equally occurs for both RCP and LCP. The negative axial-polarization of the dipolar SP, reflecting a backward evanescent field at the surface of the nanoparticle [26,27], renders electron density formed primarily near the front side of the nanoparticle where a light wave encounters first, making the conduction electrons of dipolar SPs be repelled or diffused through the surface toward the back side via a ponderomotive force [64] or as a result of an electromotive force [31,32]. Therefore, we conclude that the effective electric field of the longitudinal plasmons excited in the energy region of the dipolar SP resonance is parallel to the wavevector k (Figure 2b). Exact Mie calculations show that the field lines of the Poynting vector inside a metal nanoparticle is backward [34,65].

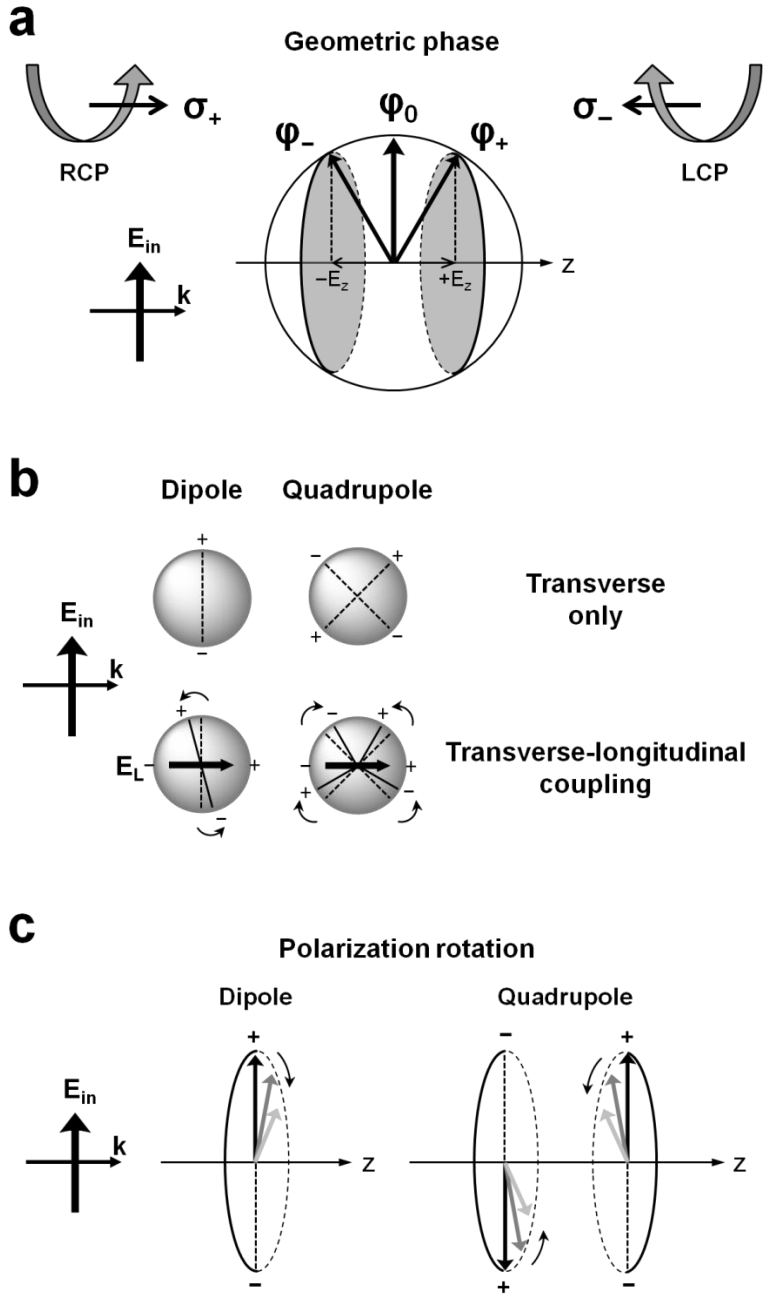

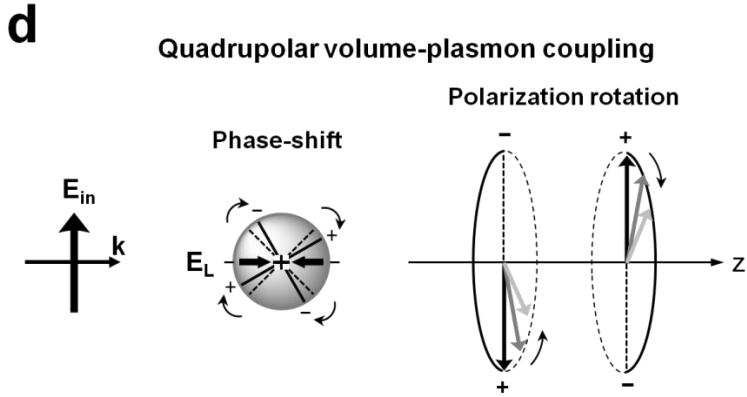

Figure 2. Determination of the absolute configuration of dipolar/quadrupolar SPs in a nanoparticle. (a) Experimentally observed geometric phase (polarization state) $\varphi_-$ of dipolar SPs represented in a Poincaré sphere. Ordinary geometric phase $\varphi_0$ is aligned with incident electric field $E_{in}$. RCP (LCP) with helicity $\sigma_+$ ($\sigma_-$) is a right (left) circularly polarized plane wave. Phase helicity is denoted by + and − signs in $\varphi$ and $\sigma$. $E_z$ is the axial component of the angular momentum of dipolar SP, parallel to the wavevector k. (b) Real-space description of coupling of longitudinal volume plasmons to surface plasmons with conduction-electron movements indicated. $E_L$ (black arrow) represents the direction of longitudinal field. Phase shift is indicated by curved arrows. For the quadrupole, the separate dipole on the left side is called a front dipole. (c) Polarization rotation for dipolar and quadrupolar SPs. Dipole rotates counter-clockwise viewed from the light source. Two separate dipoles on the quadrupole rotate in the opposite direction: clockwise for the front and counter-clockwise for the back. (d) Coupling of the quadrupolar SP to the volume plasmon of a high-order mode representing spherical standing waves. With two dipoles rotating in the same direction (clockwise), phase shift is coherent. Degrees of phase-shift and polarization rotation are exaggerated for visualization.

Two opposite electric-dipoles comprising a quadrupolar SP are phase-shifted either in a unilateral or concerted manner, depending on the direction of longitudinal electric field. The sign of the phase-shift for a separate dipole can be determined from conduction-electron movements compared to those in the phase-shifted dipolar SP. Because of the retardation on a quadurpolar SP by phase $\pi$, one of the two electric dipoles on the quadrupolar SP is opposed to the dipolar SP. Given the direction of the effective longitudinal field parallel to the wavevector k across the nanoparticle, one-half of the conduction electrons move through the surface in the same direction as the dipolar SP, but the other half move in the opposite direction that results in a positive phase-shift (Figure 2b). On the basis of the experimental observation of the net positive ellipticity of quadrupolar SPs, the rotational strength of the front dipole should be stronger than that of the back dipole, consistent with the electric-field enhancement primarily on the front dipole [66]. A similar asymmetry of the electric quadrupole in metal nanoparticles was observed in the polar plot of the hyper Rayleigh scattering, exposing an angle-shifted and amplitude-varied pattern of the four quadrupolar lobes [67].

Normally, the formation of an axial component of an electric field through a phase-shift implicates polarization rotation concomitantly arisen in an electromagnetic field [68]. The vector of the dephased electric field of the dipolar SP rotates counter-clockwise when viewed from the light source toward the detector (Figure 2c), equivalent to an elliptical polarization of incident light, meaning that the usual dipolar SP is a helical mode associated with having radial and polar components [69], and the similar components describe an optical vortex beam [70,71]. On the other hand, the front dipole of the quadrupolar SP rotates in the opposite direction whereas the back dipole rotates in the same direction as the dipolar SP (Figure 2c), explicating the small molar ellipticity of the quadrupolar SP when energetically overlapped with a dipolar SP. Peculiarly, the phase-shift and polarization rotation of a quadrupolar SP can also occur in a concerted manner. Figure 2d presents a high-order mode of volume plasmons, representing spherical standing waves, which two opposite longitudinal fields point to the center of nanoparticle. In this quadrupolar coupling mode, both the phase shift (all positive) and the polarization rotation (all clockwise viewed from the light source) of the two opposite dipoles are coherent. This provides a theoretical explanation for the sum rule for rotational strength of conduction-electron oscillations observed in the 150-nm Au spheres where the quadrupolar SPs are energetically separated from the dipolar SPs (Figure 1g).

Until now, we have qualitatively described the plasmonic handedness using phase shift and polarization rotation. We formulate a rotational strength of surface plasmons by a scalar triple product using the longitudinal and transverse electric fields. The rotational strength of plasmons ($R_{plasmon}$) may be defined as follow:

$$R_{plasmon} = \lambda \cdot (E_L \times E_T)$$

where $\lambda$ is the coupling vector or the position vector of the plasmon on the nanoparticle surface, $E_L$ and $E_T$ represent the effective electric field vector of longitudinal volume plasmons and ordinary transverse surface plasmons, respectively. This is an analogue of the classical two-group electric-dipole mechanism of optical activity, $r_{21} \cdot (\mu_{02} \times \mu_{01})$ where r denotes the position vector of the charge [1,16,17]. The angle between $\lambda$ and $E_L \times E_T$ is > 90° for a negative rotational strength and < 90° for a positive rotational strength. A magnetic field must involve the polarization rotation ($\lambda \times E_L$), but it can be trivially determined by the sign of phase shift $E_L \times E_T$ that has a form of angular momentum. Note that $E_L$ breaks time-reversal symmetry. A dipolar SP has a negative $E_L \times E_T$ (i.e., left-handed torque) while a quadrupolar SP has a positive torque (i.e., right-handed torque), and the observed sum rule for rotational strength of SPs can be considered to be the conservation of angular momentum in SPs. On the contrary to the rotational strength based on the quantum mechanical model of optical activity R = Im{μ·m}, which geometrical chirality is described by, $R_{plasmon}$ does not require chiral morphology, and it can be applied to planar metal nanostructures where strength of the two electric fields is substantial and a magnetic field is of minor significance.

## IV. CD IN THE INTERBAND TRANSITION

In the near-UV region where d-band electrons are excited to sp-bands near the Fermi level, only positive values of ellipticity are observed (Figure 1h). The ellipticity values were obtained at a wavelength of 350 nm to avoid spectral overlap with surface gold complexes. The amplitude of the positive ellipticity, which becomes measurable at the diameter larger than 20 nm, increases as nanoparticle size increases, implying that the CD in the interband transition is a bulk property rather than a surface anisotropy effect [40]. Since high-order modes of volume plasmons also occur in the UV region [27,35] and they generate inhomogeneous internal electric-fields [30], CD in the interband transitions is attributed to volume plasmons. In particular, the resulting field gradient splits the d-band into parity-even and parity-odd modes. Only the parity-odd modes respond to CD, which this opto-magnetic phenomenon is similar to the inverse Faraday rotation [46], resulting in the positive rotational strength for the optical excitation of bound d-electrons that a Lorentz force acts on. The spatial-dispersion effect generally becomes more pronounced as the wavelength of light decreases [22], indicated by the gradually increasing CD amplitude in the UV region. A detailed description will be reported elsewhere [72].

## V. EFFECTS OF SHAPE, d-BAND, AND METAL TYPE

Volume plasmons produce the size-dependent CD in the SP resonances of single nanocrystals, irrelevant to crystal shape, d-electron contribution, or metal type, concluded by the following experimental observations. First, we examined whether or not the subtle irregularity of the crystal structure for the quasi-spherical Au nanoparticles [67] is considerable in the observed CD response by comparing with CD responses of high-symmetry nanoparticles. Figure 3 shows CD and UV-vis extinction spectra of triangular and octahedral Au nanoparticles. The Au triangles with edge length of 55 nm and thickness of about 10 nm exhibit negative ellipticities of the in-plane dipolar SPs at the wavelength around 620 nm (Figure 3a), whereas the Au octahedra with edge length of 103 nm exhibit positive ellipticities centered at the plasmon wavelength of 600 nm (Figure 3b). Those symmetric nanoparticles show CD response, and this confirms that the irregular shape of quasi-spherical nanoparticles is not the source of the observed CD as any signal arisen by the shape irregularity is origin-dependent and must be averaged to zero in solution. Second, the negative ellipticity of the Au triangles confirms the CD in the SPs must be the sole property of conduction-electrons, because of the interband transition edge located at around 520 nm in gold [10], not a partial contribution from d-electrons that often accounts for the optical Kerr effect [73].

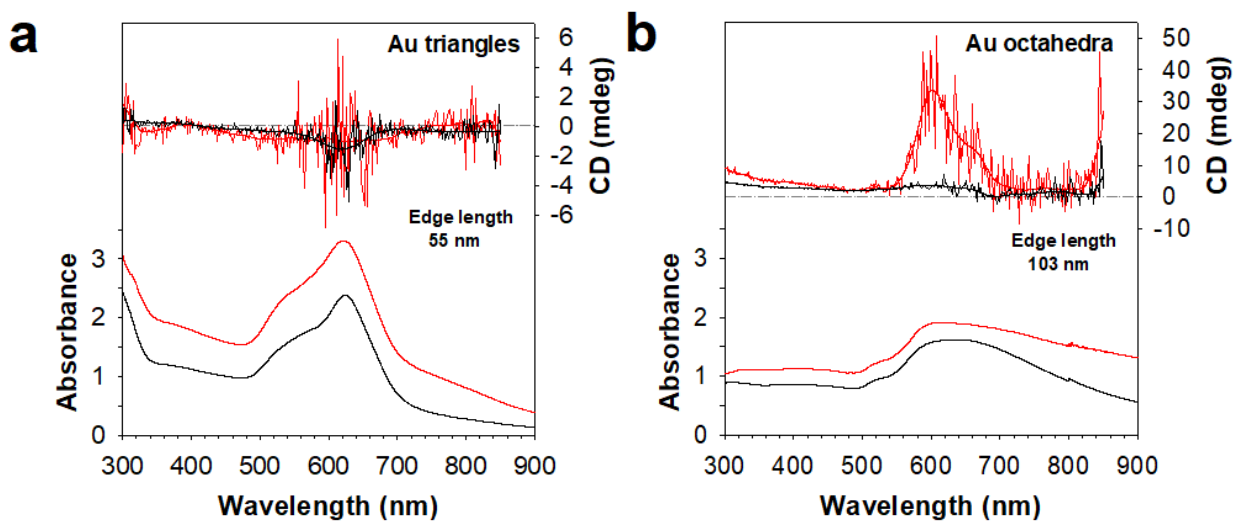

Figure 3. CD and UV-vis extinction spectra of Au nanoparticles of high symmetry. (a) triangles;

edge length: 55 nm ±6 nm, thickness: ~10 nm, (b) octahedra; edge length: 103 nm ±18 nm. Path length: 1 cm.

Last, we examined CD responses of other noble metal nanoparticles: copper and silver. Figure 4 shows CD and UV-vis spectra of copper spheres with an average diameter of 56 nm. The Cu spheres exhibit a band of negative ellipticity in the region of plasmon wavelength around 580 nm, similar to the spherical gold nanoparticles of small sizes less than 60 nm, and gradually increasing positive ellipticities in the region of the interband transitions at 300-470 nm.

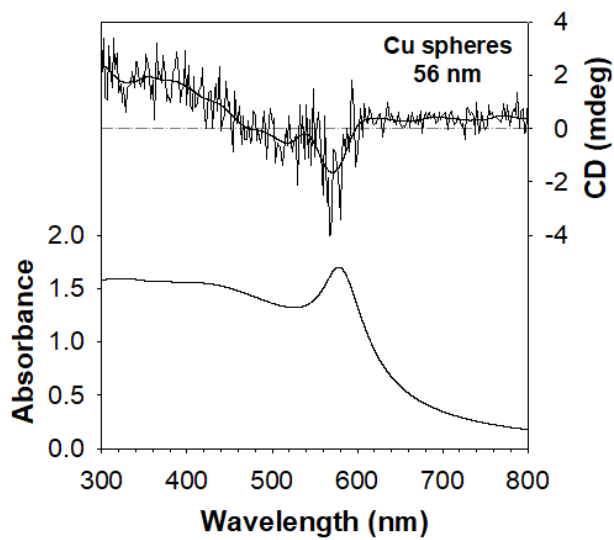

Figure 4. CD and UV-vis extinction spectra of Cu spheres. Diameter: 56 nm ±18 nm. Path length: 1 cm.

Figure 5 presents CD and UV-vis extinction spectra of silver spheres and triangles. Similar to the Au spheres, Ag spheres in the diameter range of 10-80 nm produce the size-dependent CD response in the plasmon wavelength regions between 391-449 nm (Figures 5a and 5b). However, unique aspects exist. Positive ellipticities appear at much smaller sizes for Ag, i.e., 10-20 nm in diameter, consistent with a recent report that multipolar plasmons of Ag spheres occur as small as 8 nm in diameter [74]. The similar sum rule for the rotational strength is observed for 100-nm Ag spheres, displaying a negative band of dipolar SPs excited around 490 nm and a positive band of absorptive quadrupolar SPs excited around 405 nm, although the area ratio of the two bands is deviated from unity (Figure 5c; see Methods). The larger area of the negative band reflects that the rotational strength of dipolar SPs is stronger than that of quadrupolar SPs. Interestingly, the isosbestic line at 460-500 nm on the high energy side of the dipolar SP indicates the presence not only of the two SP modes, but also probably of a plasmonic transition state having a dipole-quadrupole hybrid character in the individual nanoparticle. Molar ellipticities of two SP bands ($\pm 10^{13}$ mdeg·M$^{-1}$·m$^{-1}$) are of the same order of magnitude as those of the 150-nm Au spheres [40]. Moreover, Ag triangles with edge length of 29 nm and 37 nm exhibit only negative ellipticities of the in-plane dipolar SPs around 600 nm [75], underlining the negative rotational strength of dipolar SPs (Figure 5d).

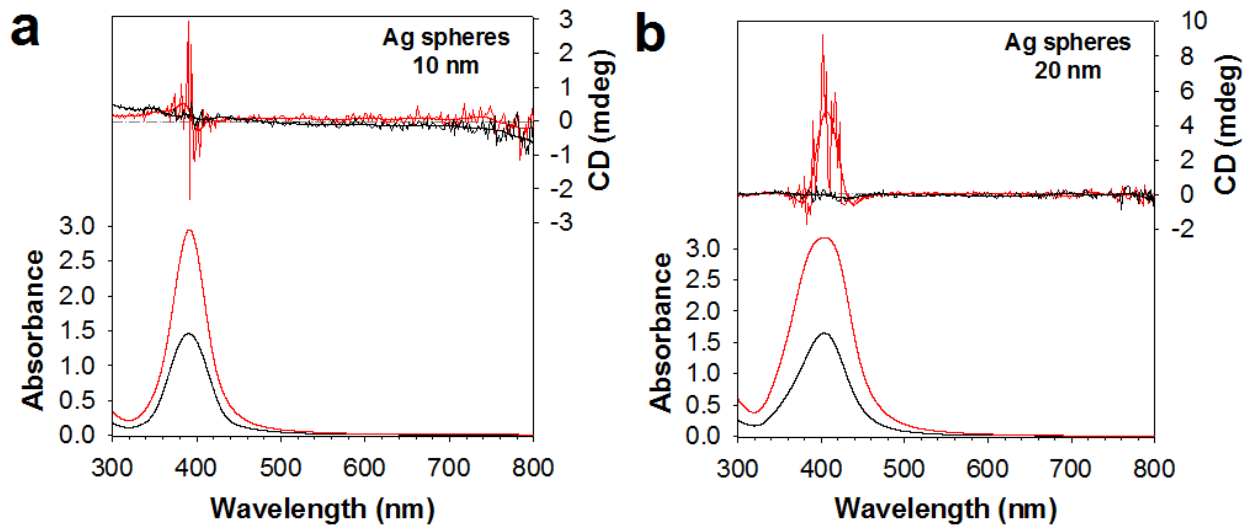

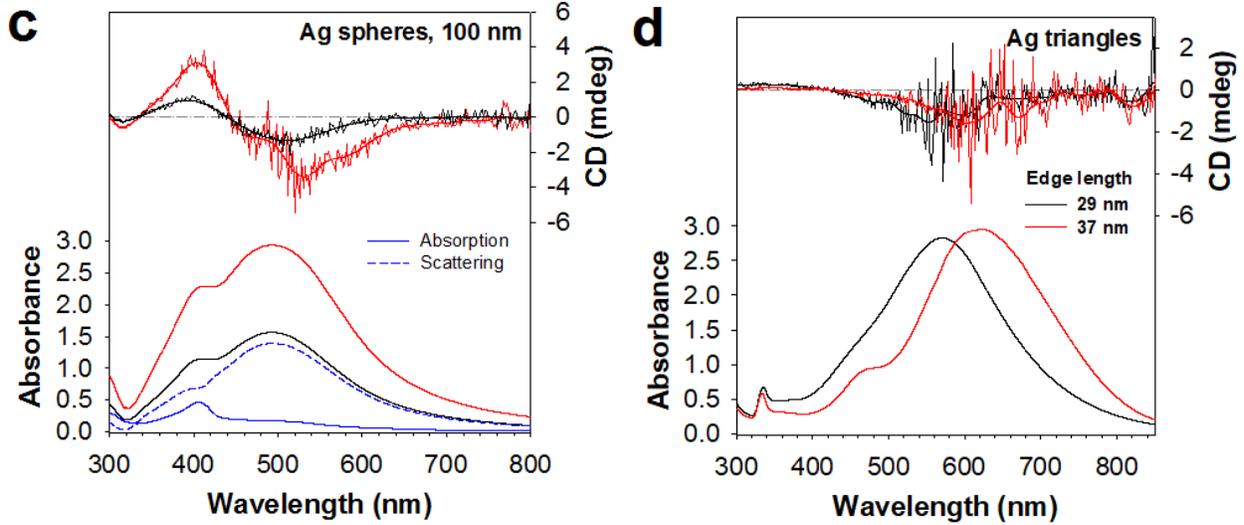

Figure 5. CD and UV-vis extinction spectra of Ag spheres and triangles. Spheres; (a) 10 nm, (b) 20 nm, (c) 100 nm in diameter. Deconvoluted absorption and scattering spectrum are also presented. Spectra of other sizes including 40, 60, and 80 nm are presented in the Supplemental Material. (d) triangles; edge lengths: 29 nm ±6 nm (black spectrum), 37 nm ±7 nm (red spectrum), thickness < 10 nm. Path length: 1 cm.

## VI. DISCUSSION AND CONCLUSION

We may be able to qualitatively assess the rotational strength of surface plasmons (SPs) by estimating the strength of the electromagnetic pressure or optical force [31,32] that simultaneously excites longitudinal volume plasmons. The optical force acting on a dipolar SP usually becomes strong as absorption cross-section increases, where the electromagnetic pressure on the nanoparticle also becomes high. For a nanoparticle of anisotropic shape, this force is orientation- and wavelength-dependent. As the shape of a nanoparticle become elongated, this force may diminish and the rotational strength of dipolar SPs will decrease significantly when the elongated nanoparticle is oriented parallel to the wavevector. A detailed study of an aspect-ratio-dependent CD response of dipolar SPs in nanorods is a subject of ongoing work.

While experimental nonlocal response in metal nanoparticles is usually probed by a small shift of resonance energy in theoretical calculation, optical extinction spectroscopy or electron energy-loss spectroscopy [10,18,21,22,23,24,25,27], nonlocality inherent in natural optical activity is primarily manifested by CD spectroscopy. Our CD measurement results in notable molar ellipticity of dipolar SPs ranging -5 × ($10^8$-$10^{11}$) mdeg·$M^{-1}$·$m^{-1}$ for spherical gold nanoparticles as sphere diameter increases from 5 to 55 nm. Dipolar SPs are optically left-handed whereas quadrupolar SPs are optically right-handed, and the plasmonic handedness can be described by the scalar vector product $\lambda \cdot (E_L \times E_T)$ where $\lambda$ is the coupling vector. Volume-plasmon-induced CD may be considered as an optical Kerr effect which only conduction-electrons have the effect on. Longitudinal volume plasmons that are optically excited simultaneously with SPs in individual metal nanoparticles bring about the same effect as electrogyration [76] (i.e., the field-induced CD) so that SPs can be said to be intrinsically chiral, although structural dissymmetry is not present. No relation was found between structure and CD in the conduction-electron oscillations, enabling for exclusion of other effects, such as anisotropic assembly of nanoparticles, irregular shape of individual nanoparticles, electron spin-polarization (Discussion S3 [40]), and plasmon-molecule interactions including effects of adsorbate chirality, surface charge, and solvent (Discussion S4 [40]). This study demonstrates a symmetry-breaking (parity-violation [1,2,3]) in isotropic metal nanoparticles, suggesting potential applications of single metal nanoparticles in metamaterials without need of nanofabrication and providing directional information on near-field electron dynamics and surface charge distribution for nonlinear optical response [14,41,43,44,77] or hot-electron mediated photocatalysis. Large-sized spherical nanoparticles (e.g., 150-nm Au and 100-nm Ag) produce the plasmon-mode-dependent ellipticity with substantial intensities, which are promising in applications, such as optical switching.

**ACKNOWLEDGEMENTS**

I thank Björn M. Reinhard for helpful discussion and generously allowing to use laboratory equipments, chemical reagents, and nanoparticle stock-solutions for conducting experiments; Wonmi Ahn for SEM measurements; Francesco Floris of the University of Cagliari for discussion about electronic surface states. CD spectra were collected from the CD spectrometer at the Chemical Instrumentation Center.

**APPENDIX: METHODS**

**1. Circular dichroism spectroscopy.** CD spectra were recorded using a Chirascan CD spectrometer (Applied Photophysics, Leatherhead, UK; operating range 163-900 nm). Temperature on the sample holder was maintained at 20 °C. A wavelength interval was 2 nm (step size), and acquisition time per point was 0.5 second. For each nanoparticle sample, seven spectra were subsequently collected and averaged. Water was used as the background solution, and three spectra of water have been collected at every time when the spectra of a sample were obtained, and then averaged. The averaged background spectrum was smoothened by the Savitsky-Golay algorithm. The smoothened background spectrum was subtracted from the averaged sample spectrum, and finally the resulting raw spectrum of a sample was smoothened. The averaging, subtraction, and smoothing procedures were performed using the built-in Chirascan software. Smoothened spectra were used in measuring band area and calculating molar ellipticity. Measured areas of CD bands in the spectra of 150-nm Au sphere sample of absorbance 3.0 are 236 in the 486-622 nm wavelength region and 262 in the 622-830 nm wavelength region. Those areas for 100-nm Ag sphere sample are 186 at 336-444 nm and 409 at 444-800 nm, yielding the ratio of 2.2 (2.4 for the sample of absorbance 1.5). Ellipticity values for the interband transition at a wavelength of 350 nm were selected from spectra of the Au-sphere sample of absorbance 3.0. Since values of absorbance at 350 nm ranges 1.6-2.3, ellipticity values were normalized relative to absorbance of 1.7. The normalized values of ellipticity are presented in the plot, but the values for 5- and 10-nm Au spheres are omitted because of a high noise level in the near-UV region. The choice of a cuvette is important to measure CD spectra and to avoid obtaining an artifact CD spectrum. A fluorescence cell (Hellma GmbH & Co. KG, model 111-10-40 QS, light path 10 mm, width 10 mm) was used in this study. A CD signal of the individual nanoparticles is not strong enough to be detected with a standard 1.0-mm cuvette employed. A cuvette having a smaller width than a beam diameter, including a micro-volume cell and a 4-mm inner-width cell, usually generates an artifact birefringence. Whether or not a cuvette produces an artifact spectrum is readily verified by monitoring spectral reproducibility while placing a cuvette with absorptive solution contained in the opposite direction.

**2. UV-vis extinction spectroscopy.** Extinction spectra of nanoparticle solutions were collected using a Cary 5000 UV-Vis-NIR spectrophotometer (Varian, Palo Alto, CA). Deconvolution of a spectrum to absorption and scattering spectra was performed using a Mie theory calculator.

**3. Nanocrystal sample preparation.** Aqueous solutions of spherical gold and silver nanoparticles stabilized by citrate were obtained from BBI Solutions, Ted Pella, Nanopartz, or nanoComposix. Gold triangles, gold octrahedra, copper spheres, and silver triangles were synthesized using cetyl trimethylammonium chloride (CTAC), polyvinylpyrrolidone (PVP), poly(acrylic acid), and citrate/PVP, respectively (see Supplemental Material). For CD and UV-vis extinction measurements, all nanoparticle solutions were centrifuged once and sedimented nanoparticles were redispersed in water that was purified by a Barnstead Nanopure system. For each sample of nanoparticles, typically two solutions of different concentration were prepared:

absorbance of about 1.5 and 3.0 at the wavelength of plasmon resonance. Extent of nanoparticle aggregation is minor indicated by the negligible spectral red-shift in UV-vis spectra even at these high concentrations. To study plasmon-molecule interactions, spherical gold nanoparticles were synthesized or functionalized using other molecules, including D(L)-tryptophan, L-ascorbic acid, tannic acid, mercaptopropionic acid, ethanol, citrate, acrylate, polyethyleneimine, and PVP (see Supplemental Material).

**4. Nanocrystal size and shape characterization.** Average diameter values of the commercially available nanoparticles provided by the suppliers were taken without extensive size analysis, and nominal coefficients of variation are typically ≤ 8% for gold (< 6% for 32, 47, 55, 65, 69 and 88 nm; ≤ 10% for 10 nm; ≤ 15% for 5 nm) provided by BBI Solutions or Nanopartz, and < 15% for silver (< 25% for 10 nm) provided by Ted Pella or nanoComposix. In this study, these average sizes were briefly verified by scanning electron microscopy (SEM), dynamic light scattering (DLS), and the well-known size-dependence of plasmon wavelengths. Sizes and shapes of the synthesized nanoparticles were analyzed by SEM, typically with about 100 particles examined, as well as by DLS (see Supplemental Material).

[20] M. I. Tribelsky, A. E. Miroshnichenko, and Y. S. Kivshar, *Unconventional Fano resonances in light scattering by small particles,* Europhys. Lett. **97,** 44005 (2012).

[21] S. Raza, S. I. Bozhevolnyi, M. Wubs, and N. A. Mortensen, *Nonlocal optical response in metallic nanostructures,* J. Phys. Condens. Mat. **27,** 183204 (2015).

[22] S. Raza, G. Toscano, A.-P. Jauho, M. Wubs, and N. A. Mortensen, *Unusual resonances in nanoplasmonic structures due to nonlocal response,* Phys. Rev. B **84,** 121412(R) (2011).

[23] T. V. Teperik, P. Nordlander, J. Aizpurua, and A. G. Borisov, *Robust Subnanometric Plasmon Ruler by Rescaling of the Nonlocal Optical Response,* Phys. Rev. Lett. **110,** 263901 (2013).

[24] Y. Luo, A. I. Fernandez-Dominguez, A. Wiener, S. A. Maier, and J. B. Pendry, *Surface Plasmons and Nonlocality: A Simple Model,* Phys. Rev. Lett. **111,** 093901 (2013).

[25] J. M. McMahon, S. K. Gray, and G. C. Schatz, *Nonlocal Optical Response of Metal Nanostructures with Arbitrary Shape,* Phys. Rev. Lett. **103,** 097403 (2009).

[26] R. Ruppin, Spherical and cylindrical surface polaritons in solids, in *Electromagnetic Surface Modes*, edited by A. D. Boardman (John Wiley & Sons, 1982).

[27] R. Ruppin, *Optical properties of small metal spheres,* Phys. Rev. B **11,** 2871 (1975).

[28] V. B. Gildenburg, V. A. Kostin, and I. A. Pavlichenko, *Resonances of surface and volume plasmons in atomic clusters,* Phys. Plasmas **18,** 092101 (2011).

[29] P. de Andrés, R. Monreal, and F. Flores, *Quantum size and nonlocal effects in the electromagnetic properties of small metallic spheres,* Phys. Rev. B **32,** 7878 (1985).

[30] W. Jacak *et al., Radius dependent shift in surface plasmon frequency in large metallic nanospheres: Theory and experiment,* J. Appl. Phys. **107,** 124317 (2010).

[31] M. Durach, A. Rusina, and M. I. Stockman, *Giant Surface-Plasmon-Induced Drag Effect in Metal Nanowires,* Phys. Rev. Lett. **103,** 186801 (2009).

[32] M. Durach and N. Noginova, *On the nature of the plasmon drag effect,* Phys. Rev. B **93,** 161406(R) (2016).

[33] H. Xu, *Electromagnetic energy flow near nanoparticles-I: single spheres,* J. Quant. Spectrosc. Radiat. Transfer **87,** 53 (2004).

[34] Z. B. Wang, B. S. Luk'yanchuk, M. H. Hong, Y. Lin, and T. C. Chong, *Energy flow around a small particle investigated by classical Mie theory,* Phys. Rev. B **70,** 035418 (2004).

**Supplemental Material:**

# Observation of intrinsic chirality of surface plasmon resonances in single nanocrystals

(Dated: March 1, 2018)

## Supporting Methods

**Scanning electron microscopy (SEM)**. SEM images of nanoparticles were obtained by a Zeiss SUPRA 40VP scanning electron microscope. Each SEM sample was prepared with nanoparticles spread on a 7 mm × 7 mm silicon wafer. SEM images were used for nanoparticle size/shape analyses.

**Dynamic light scattering (DLS)**. Size distributions and average diameters of spherical nanoparticles were determined by DLS using a Malvern ZetaSizer.

**Zeta-potential measurements**. Zeta ($\zeta$) potential of spherical gold nanoparticles were measured using the Malvern ZetaSizer.

**Materials.** Spherical gold nanoparticles were purchased from the following suppliers: 5, 10, 20, 60, 80, and 150 nm from BBI Solutions (cv.$\leq$ 8% for 20, 60, 80, and 150 nm; $\leq$ 10% for 10 nm; $\leq$ 15% for 5 nm), 32, 47, 55, 65, 69, and 88 nm from Nanopartz (cv.< 6%). Spherical silver nanoparticles were purchased from the following suppliers: 10 nm from nanoComposix (cv.< 25%), 20, 40, 60, 80, and 100 nm from Ted Pella (cv.< 15%). The diameter values are within ± 1 nm for all sizes of gold and silver nanoparticles. All chemicals were used without further purification. Gold(III) chloride trihydrate ($HAuCl_4 \cdot 3H_2O$, >99.9%), copper(II) sulfate pentahydrate ($CuSO_4 \cdot 5H_2O$, 99.995%), silver nitrate ($AgNO_3$, $\geq$99.0%), polyvinylpyrrolidone (PVP, $M_w$ 55000), 1,5-pentanediol (96%), cetyl trimethylammonium bromide (CTAB, 99%), sodium citrate tribasic dihydrate ($Na_3Citrate \cdot 2H_2O$,

≥99.0%), cetyl trimethylammonium chloride (CTAC, >95.0%; TCI CO., LTD), potassium iodide (≥99.5%), L-ascorbic acid (>99.5%; Fluka), sodium hydroxide (NaOH, ≥98%), poly(acrylic acid) ($M_w$ 1800), hydrazine hydrate solution (78 – 82%), PVP K15 ($M_w$ 15k Da, $M_r$ ~10,000; Fluka), sodium borohydride ($NaBH_4$, 99.99%), hydrogen peroxide solution (30 wt.% in water, 9.8 M), D-tryptophan (≥98%), L-tryptophan (≥98%), tannic acid, sodium acrylate (97%), polyethyleneimine (PEI, 50 wt% solution in water, $M_w$ 750,000, $M_n$ 60,000), 3-meracaptopropionic acid (≥99%), and ethanol (200 proof) were obtained from aforementioned suppliers or Sigma-Aldrich unless stated. Water was purified by a Barnstead Nanopure system (17.8 MΩ/cm). All glassware was washed with aqua-regia (3:1, $HCl/HNO_3$). *Caution! The aqua-regia solution are highly corrosive. It should be handled with extreme care and appropriate safety precautions.* The aqua-regia cleaned glassware was rinsed thoroughly with water.

**Preparations of Au, Cu, Ag nanocrystals**

Purified water was used as the solvent unless otherwise stated.

**Au triangles.** Au triangles were synthesized according to a method in the literature.[1] 1.6 mL of 0.1 M CTAC solution was added in a 20-mL glass vial with 3.7 mL water contained, and then the following solutions were subsequently added: 75 μL of 0.01 M potassium iodide, 4.37 mL of 0.465 mM $HAuCl_4$ (stock solution), and 20.3 μL of 0.1 M NaOH. 100 μL of 0.064 M ascorbic acid was added to the resulting mixture while moderately shaking the vial. Finally, 10 μL of 0.1 M NaOH solution was added immediately with the vial shaken for a few seconds. Average edge length measured by SEM: 55 nm ± 6 nm.

**Au octahedra.** Au octahedra were synthesized according to a method in the literature.[2] 0.00425 g of $AgNO_3$ was dissolved in 10 mL of pentanediol, and 0.10 mL of the resulting $AgNO_3$ solution was added to 5 mL of boiling pentanediol (b.p. 242 °C) in a glass vial while stirring. 0.223 of PVP ($M_w$ 55000) was dissolved in 3 mL of pentanediol, and 0.059 g of $HAuCl_4·3H_2O$ was dissolved in 3 mL of pentanediol. Aliquots of the resulting PVP and of $HAuCl_4$ solutions were alternatively added to the boiling $AgNO_3$ solution by 200 µL every 30 seconds. This resulted in the final molar ratio of $Ag^+/Au^{3+}$ = 1/595. The mixture solution was heated at around 235 °C for an hour while stirring. After cooled down, the nanoparticle solution was purified by centrifugation/dispersion with using ethanol, and then the purified nanoparticles were stored in ethanol. Average edge length measured by SEM: 103 nm ± 18 nm.

**Cu spheres.** Spherical copper nanoparticles were synthesized according to a method in the literature.[3] Both 0.0125 g of $CuSO_4·5H_2O$ and 0.025 g of poly(acrylic acid) (PAA, $M_w$ 1800) were dissolved together in 10 mL water, and the resulting concentration of PAA was 1.33 mM. The mixture solution was stirred at around 60 °C. After 20 minutes, pH of the solution was adjusted to 10.0 by adding dropwise 0.5 M NaOH, and then 10 minutes later 1.28 uL of the hydrazine reagent that was diluted with 1 mL of water was added to the $CuSO_4$/PAA solution. The subsequent reaction was stopped typically within 30 minutes before bluish color in the solution appeared. Average diameter measured by SEM: 56 nm ± 18 nm.

**Ag triangles.** Ag triangles were synthesized according to a method in the literature.[4] In 12 mL of water, 0.1 mL of 0.0125 M $AgNO_3$, 0.25 mL of 75 mM $Na_3$Citrate, 0.1 mL of 17.5 mM PVP 15K, and 30 µL of the hydrogen peroxide reagent were added and vigorously stirred. Four samples of this solution were prepared, and varied amounts of 100 mM $NaBH_4$ solution, i.e., 100, 125, 150, 300 µL, were added to each sample. The 100-µL (150-µL) $NaBH_4$ sample produced Ag triangles with average edge length of 37 nm ± 7 nm (29 nm ± 6 nm), respectively, which were determined by SEM.

**Au spheres stabilized by tryptophan (Trp).** Trp-stabilized Au spheres were synthesized according to a method in the literature.[5] 0.8 mL of 1 mM $HAuCl_4$ solution and 1 mL of 1 mM D-Trp or L-Trp were mixed and then heated at around 50 °C in the dark overnight. Average diameter measured by SEM: D-Trp-Au 60 nm ± 7 nm, L-Trp-Au 61 nm ± 7 nm.

**Au spheres stabilized by L-ascorbic acid.** L-ascorbic acid (AA) stabilized Au spheres were synthesized according to a method of synthesizing core-shell nanoparticles in the literature.[6] First, in order to prepare seed nanoparticles, citrate-stabilized Au spheres with an average diameter of 18 nm were synthesized by the Turkevich method,[7] which 8.2 mL of 10 mM $Na_3$Citrate (stock solution) was added to 100 mL of 0.30 mM HAuCl$_4$ solution while boiling. 5 mL of the resulting Au sphere solution was diluted to 25 mL. 5 mL of 5 mM (0.2 wt.%) HAuCl$_4$·3H$_2$O solution was diluted to 25 mL. 1.25 mL of 56.8 mM (1 wt.%) AA solution and 0.6125 mL of 34 mM (1 wt.%) Na$_3$Citrate·2H$_2$O solution were mixed and diluted to 25 mL. 2.7 mL of the HAuCl$_4$ solution and 2.7 mL of the AA/Na$_3$Citrate solution were added dropwise to the diluted seed solution while stirring, resulting in growth of the 18-nm seeds to 25-nm spheres stabilized by AA, measured by DLS ($\lambda_{max}$ in the UV-vis extinction spectrum: 521 nm).

**Au spheres stabilized by tannic acid.** Tannic acid stabilized Au spheres were synthesized according to a method in the literature.[8] 10 mL of 6 mM tannic acid solution was added to a flask containing 50 mL of 0.13 mM HAuCl$_4$ solution while boiling. After 30 minutes, the flask was removed from heating and cooled down at room temperature, yielding about 29-nm spheres measured by SEM and DLS ($\lambda_{max}$ in the UV-vis extinction spectrum: 527 nm).

**Ag spheres stabilized by tannic acid.** Aqueous solution of tannic acid stabilized Ag spheres with an average diameter of 20 nm was obtained from Ted Pella. Average diameter verified by SEM: 19 nm ± 3 nm.

**Au spheres stabilized by citrate.** Citrate stabilized Au spheres were synthesized by the Turkevich method.[7] Briefly, 4.0 mL of 10 mM Na$_3$Citrate stock solution was added to 100 mL of 0.30 mM HAuCl$_4$ solution while boiling, yielding about 47-nm spheres estimated by SEM and the plasmon wavelength ($\lambda_{max}$) of 531 nm in the UV-vis extinction spectrum (DLS mean peak: 49 nm ± 15 nm, DLS z-average: 26 nm).

**Au spheres stabilized by acrylate.** Acrylate stabilized Au spheres were synthesized according to a method in the literature.[9] 0.0098 g of HAuCl$_4$·3H$_2$O was dissolved in 40 mL of water and the resulting solution was heated to boiling. 0.03 g of sodium acrylate was dissolved in 10 mL of water and heated at 50 – 60 °C, and then added to the HAuCl$_4$ solution. The final concentrations of HAuCl$_4$ and sodium acrylate were 0.5 mM and 6.38 mM, respectively. The mixture solution was being boiled for 30 minutes, yielding about 36-nm spheres measured by SEM and DLS ($\lambda_{max}$ in the UV-vis extinction spectrum: 528 nm).

**Au spheres stabilized by polyethyleneimine (PEI).** PEI stabilized Au spheres were synthesized according to a method in the literature.[10] 100 μL of 1 wt.% $HAuCl_4 \cdot 3H_2O$ solution was added to 5 mL of water. 0.41 g of the PEI reagent (50 wt.% solution in water, $M_w$ 750,000) was mixed with 5 mL of water. Both solutions were mixed and heated at around 70 °C for a day, yielding about 38-nm spheres measured by SEM and DLS ($\lambda_{max}$ in the UV-vis extinction spectrum: 530 nm).

**Au spheres stabilized by polyvinylpyrrolidone (PVP).** PVP stabilized Au spheres were synthesized according to a method in the literature.[11] 0.6 mL of 0.1 M $NaBH_4$ solution and 5 mL of 1 wt.% PVP 15K solution were added to 20 mL water in an ice bath. Aliquots of 5 mL of 10 mM $HAuCl_4$ solution and 5 mL of 1 wt.% PVP 15K solution were added to the $NaBH_4$/PVP solution by 0.25 mL every 30 seconds while stirring. The resulting solution was heated at around 80 °C for 2 hours. Average diameter determined by DLS: 12 nm.

**Au spheres functionalized with thiol.** The solution of citrate-stabilized Au spheres synthesized above (~47 nm in diameter based on the plasmon wavelength, $\lambda_{max}$, of 531 nm) was centrifuged and the sedimented nanparticles were dispersed in 0.1 mM mercaptopropionic acid solution in water.

**Au spheres in ethanol (EtOH).** The solution of citrate-stabilized Au spheres synthesized above (~47 nm in diameter based on the plasmon wavelength, $\lambda_{max}$, of 531 nm) was centrifuged and the sedimented nanparticles were dispersed in ethanol.

## Supporting Figures, Tables, and Discussions

Figures S1 – S16.

Tables S1 – S3.

Discussions S1 – S6.

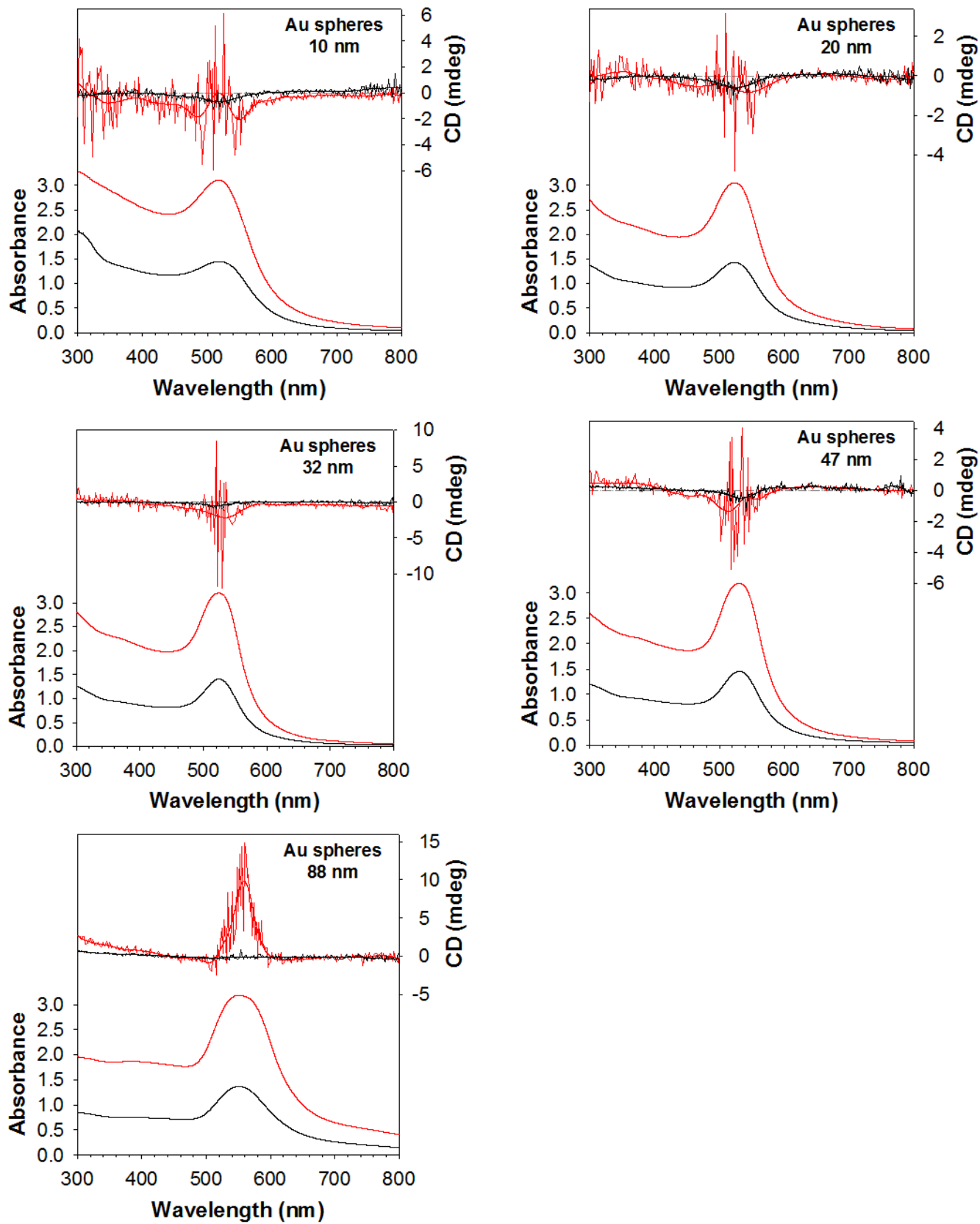

Figure S1. CD and UV-vis extinction spectra of Au spheres (10, 20, 32, 47, and 88 nm in diameter). Path length: 1 cm.

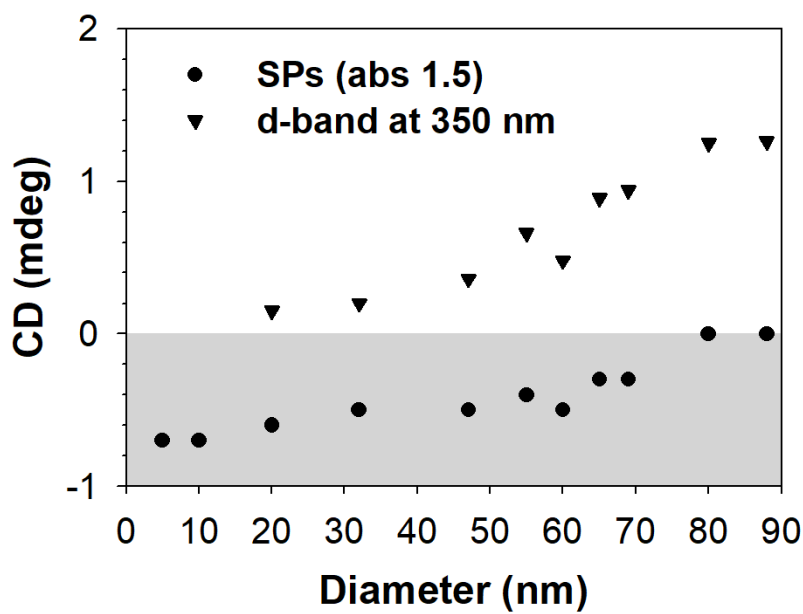

Figure S2. Ellipticity values measured at surface plasmons (SPs) and interband transitions (d-band) at a wavelength of 350 nm in Au spheres with diameter ranges of 5 – 88 nm. Plasmon wavelengths range 515 – 555 nm. Nanoparticle samples of absorbance 1.5 (3.0) at the plasmon wavelength were used for the SPs (d-band), respectively. Absorbance values at 350 nm range 1.6 – 2.3, and ellipticity values were normalized relative to absorbance of 1.7. CD data of SPs for 5 and 10-nm Au spheres are omitted because of a high noise level although their ellipticity values are presumably zero.

**Table S1. Molar ellipticities of 5 – 55 nm Au spheres and of 150-nm Au spheres at the wavelengths of dipolar SPs**

| Diameter (nm) | Absorbance | Molarity (mole/L) | Ellipticity (mdeg) | Molar ellipticity* |
|---|---|---|---|---|
| 5 | 1.47 | $1.23 \times 10^{-7}$ | -0.7 | $-5.7 \times 10^{8}$ |
| 10 | 1.45 | $1.38 \times 10^{-8}$ | -0.7 | $-5.1 \times 10^{9}$ |
| 20 | 1.43 | $1.72 \times 10^{-9}$ | -0.6 | $-3.5 \times 10^{10}$ |
| 32 | 1.41 | $3.67 \times 10^{-10}$ | -0.5 | $-1.4 \times 10^{11}$ |
| 47 | 1.46 | $1.39 \times 10^{-10}$ | -0.5 | $-3.6 \times 10^{11}$ |
| 55 | 1.62 | $0.98 \times 10^{-10}$ | -0.4 | $-4.1 \times 10^{11}$ |
| 150 | 1.0 at 530 nm | $2.80 \times 10^{-12}$ | -1.0 | $-3.6 \times 10^{13}$ |

*unit: [mdeg·M$^{-1}$·m$^{-1}$]

For example, the 5-nm gold spheres exhibit:

ellipticity of (-) 0.7 mdeg when absorbance is 1.47 at the plasmon wavelength, resulting in molarity = $1.23 \times 10^{-7}$ moles/L = $1.23 \times 10^{-7}$ M, and the path-length is $10^{-2}$ m. Thus, molar ellipticity is (-) 0.7 mdeg/($1.23 \times 10^{-7}$ M × $10^{-2}$ m) = (-) $5.7 \times 10^{8}$ mdeg·$M^{-1}$·$m^{-1}$.

Molar ellipticity [θ] in millidegrees·$M^{-1}$·$m^{-1}$ is equivalent to millidegrees·$cm^{2}$·$dmol^{-1}$, where d is the path-length in cm.

**Table S2. Molar ellipticities of 5 – 88 nm gold spheres at a wavelength of 350 nm**

| Diameter (nm) | Abs at SPs | Molarity (mole/L) | Abs at 350 nm | Ellipticity (mdeg) | Normalized ellipticity for Abs 1.7 at 350 nm | Molar ellipticity at 350 nm[†] |
|---|---|---|---|---|---|---|
| 20 | 3.05 | $3.7 \times 10^{-9}$ | 2.24 | 0.20 | 0.15 | $4.05 \times 10^{9}$ |
| 32 | 3.20 | $8.3 \times 10^{-10}$ | 2.35 | 0.28 | 0.20 | $2.41 \times 10^{10}$ |
| 47 | 3.19 | $3.0 \times 10^{-10}$ | 2.18 | 0.47 | 0.36 | $1.20 \times 10^{11}$ |
| 55 | 2.91 | $1.7 \times 10^{-10}$ | 1.58 | 0.62 | 0.66 | $3.89 \times 10^{11}$ |
| 60 | 2.95 | $1.3 \times 10^{-10}$ | 1.71 | 0.49 | 0.48 | $3.69 \times 10^{11}$ |
| 65 | 2.95 | $1.0 \times 10^{-10}$ | 1.70 | 0.90 | 0.89 | $8.90 \times 10^{11}$ |
| 69 | 3.05 | $8.5 \times 10^{-11}$ | 1.69 | 0.94 | 0.94 | $1.11 \times 10^{12}$ |
| 80 | 3.09 | $5.6 \times 10^{-11}$ | 1.59 | 1.18 | 1.25 | $2.23 \times 10^{12}$ |
| 88 | 3.18 | $4.5 \times 10^{-11}$ | 1.85 | 1.38 | 1.26 | $2.80 \times 10^{12}$ |

[†]unit: [mdeg·$M^{-1}$·$m^{-1}$]

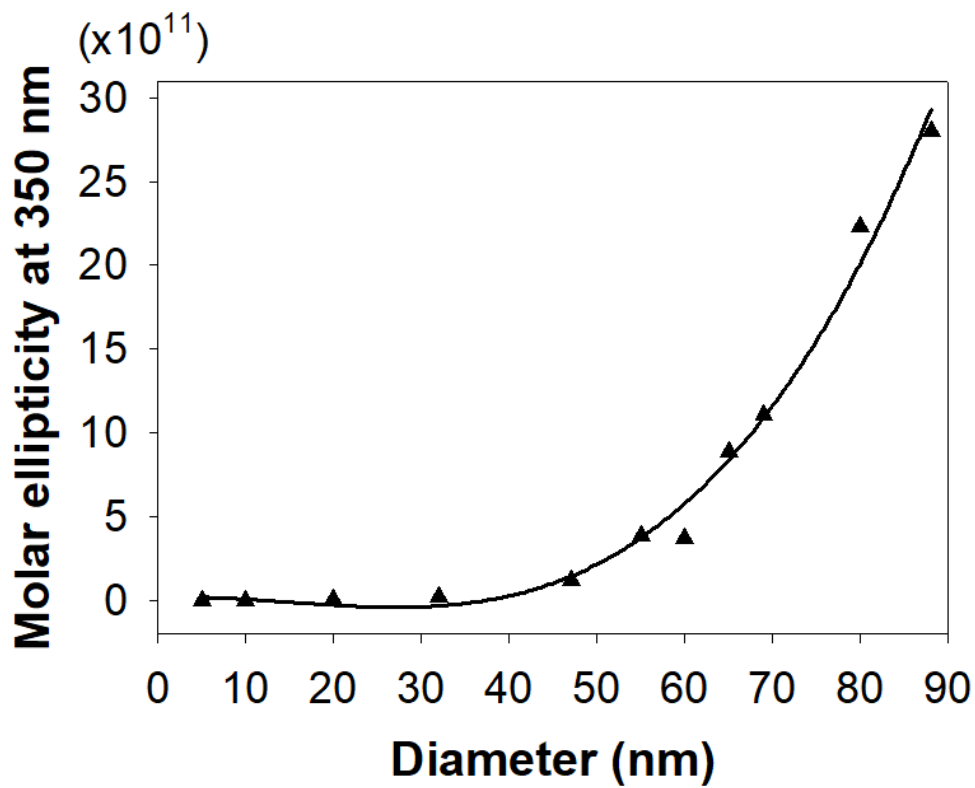

Figure S3. Molar ellipticity in the interband transition of Au sphere with diameters in the range of 5 – 88 nm measured at a wavelength of 350 nm. Ellipticity values are normalized relative to absorbance 1.7 at 350 nm. Presumably zero values of 5- and 10-nm Au spheres are included. Unit of molar ellipticity: [mdeg·M$^{-1}$·m$^{-1}$]. Values are best fit to a cubic curve, indicating the volume dependence of the CD in the interband transition. This is attributed to the volume-plasmon excitation in the near-UV region, but the spin-orbit splitting in the d-band via hybridization of $E_g$ and $T_{2g}$ d-orbital states is also volume-dependent.[12]

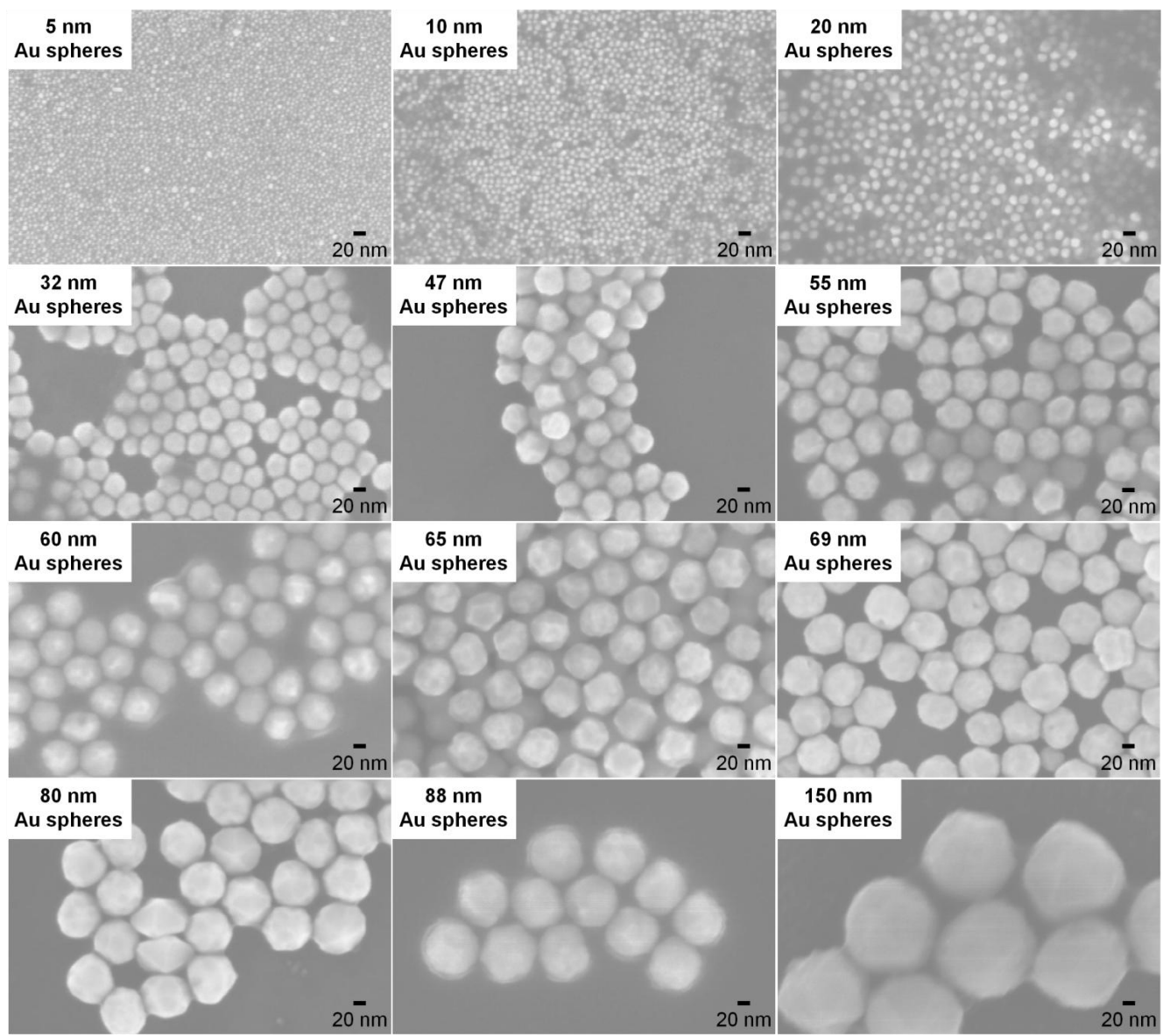

Figure S4. SEM images of Au spheres (5, 10, 20, 32, 47, 55, 60, 65, 69, 80, 88, and 150 nm in diameter).

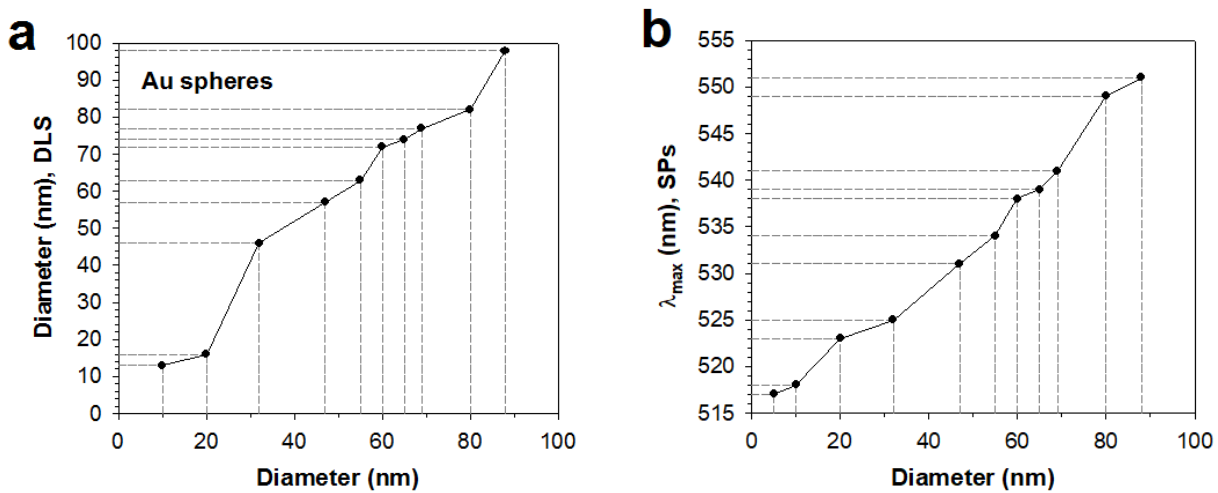

Figure S5. Average diameters of Au spheres measured by (a) DLS z-average and (b) size-dependent $\lambda_{max}$ values of SPs.

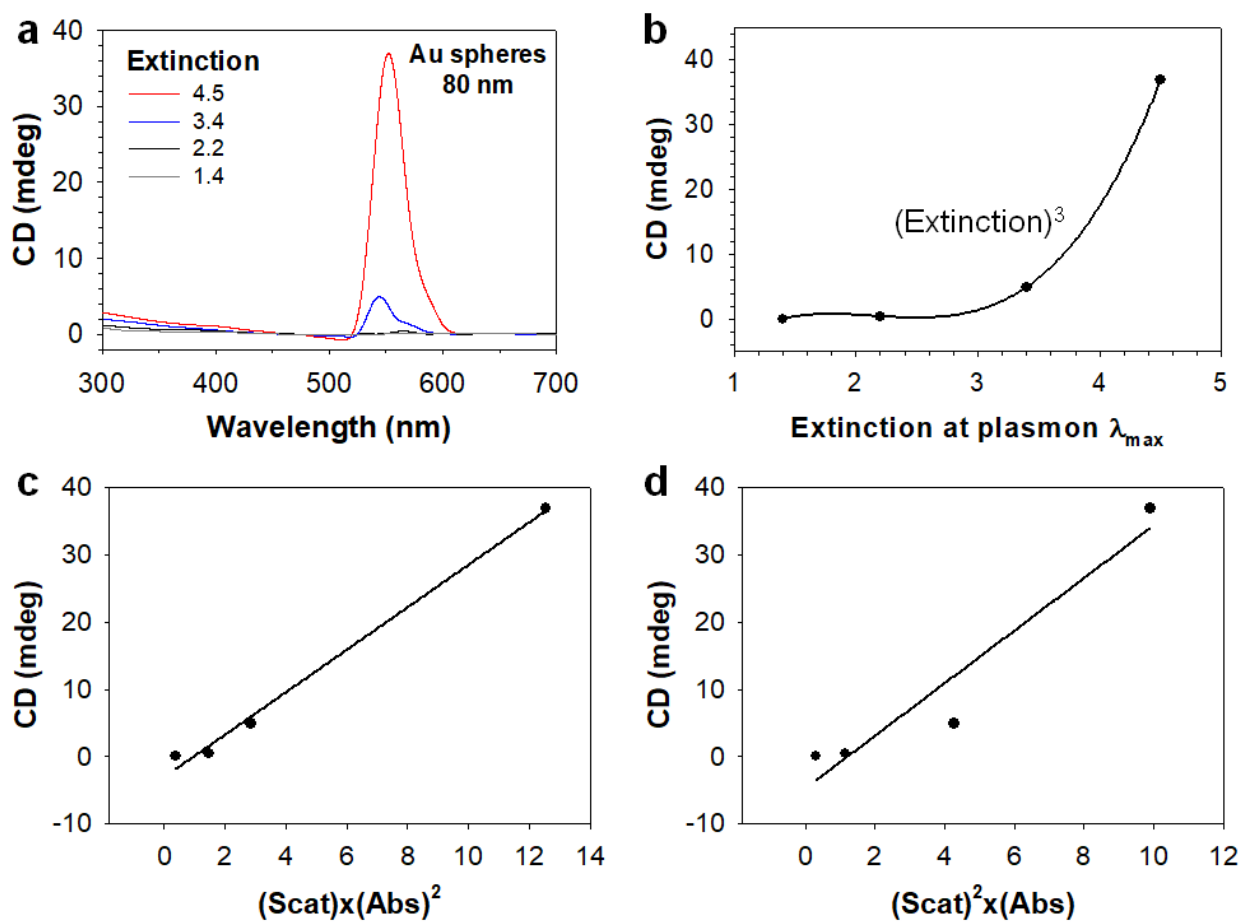

Figure S6. Amplitude modulation of the positive ellipticity of quadrupolar surface plasmons in 80-nm Au spheres. (a) CD spectra at different concentrations according to UV-vis extinction values. Path length: 1 cm. (b) CD amplitude as a function of concentration is best fit to a cubic curve. Having considered the coefficient ratio, i.e., absorption (544 nm) : scattering (562 nm) = 1.9 : 1.5, (c) the CD amplitude is linearly proportional to the product of scattering and the square of absorption, i.e., (scattering)×(absorption)$^2$ rather than (d) (scattering)$^2$×(absorption).

**Discussion S1. Spatial dispersion in crystal optics:** Spatial dispersion in crystal optics generally describes the dielectric tensor as a function of wavevector k up to the second-order:

$$\varepsilon_{ij}(\omega, \mathbf{k}) = \varepsilon_{ij}(\omega) + i\gamma_{ijl}(\omega)k_l + \alpha_{ijlm}(\omega)k_l k_m$$

where the third rank tensor (the antisymmetric tensor of gyrotropy) $\gamma_{ijl}$ vanishes in the presence of inversion symmetry.[13] The first-order term $\gamma_{ijl}$, similarly describing the effect of piezo-optical birefringence, renders natural optical activity in crystals.

**Discussion S2. Spatial dispersion in optics:** The trajectory of circularly polarized light through a high N.A. lens depends on light helicity.[14] A tightly focused beam generates a longitudinal electric field with having optical vortex.[15] The vorticity is also related to chirality,[16,17,18] although surface plasmon optical vortices have been considered only in the context of mechanical forces.[19] Experimental results demonstrates that the intensity of the longitudinal electric field is not negligible in the focal region.[20,21]

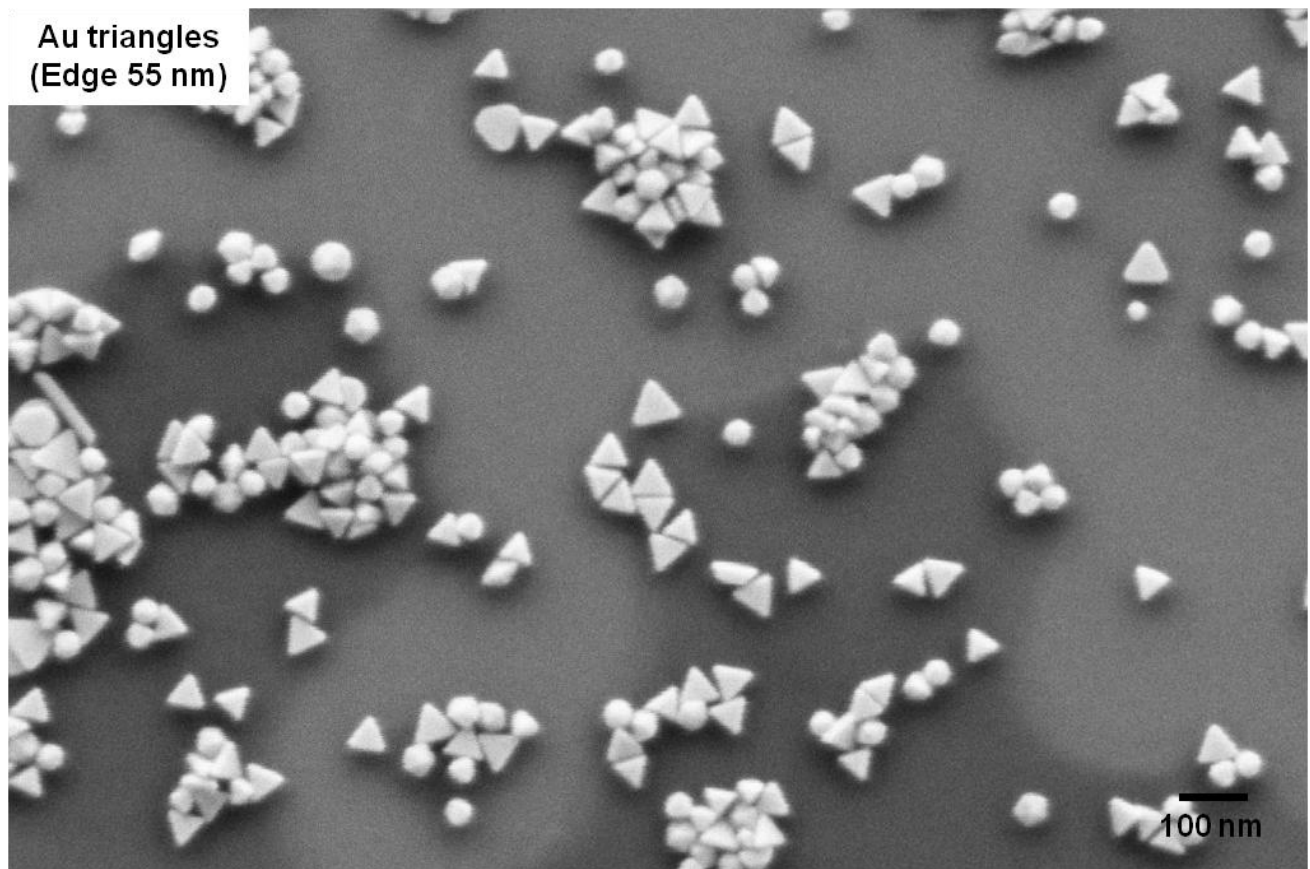
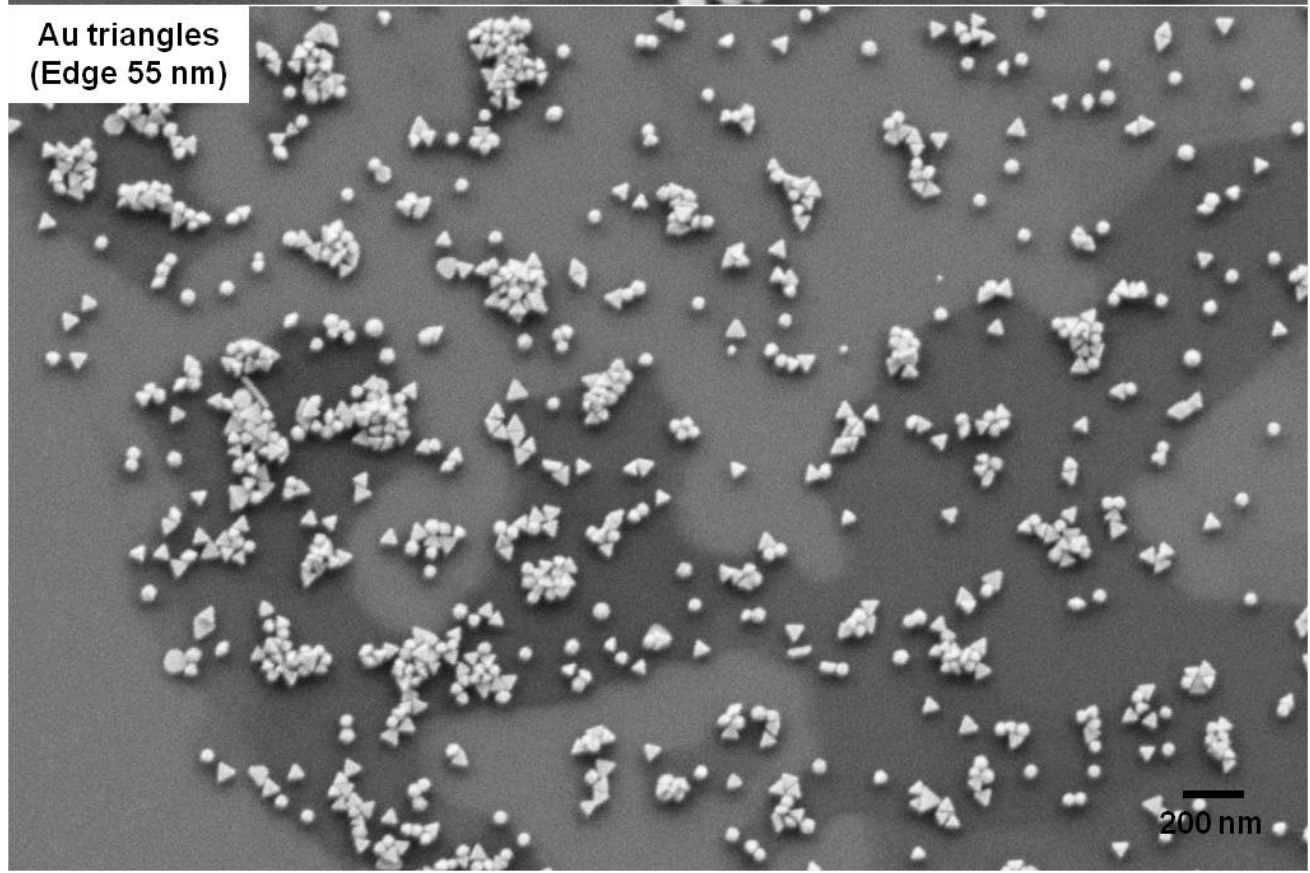

Figure S7. (a) SEM images of Au triangles. Average edge length: 55 nm ± 6 nm.

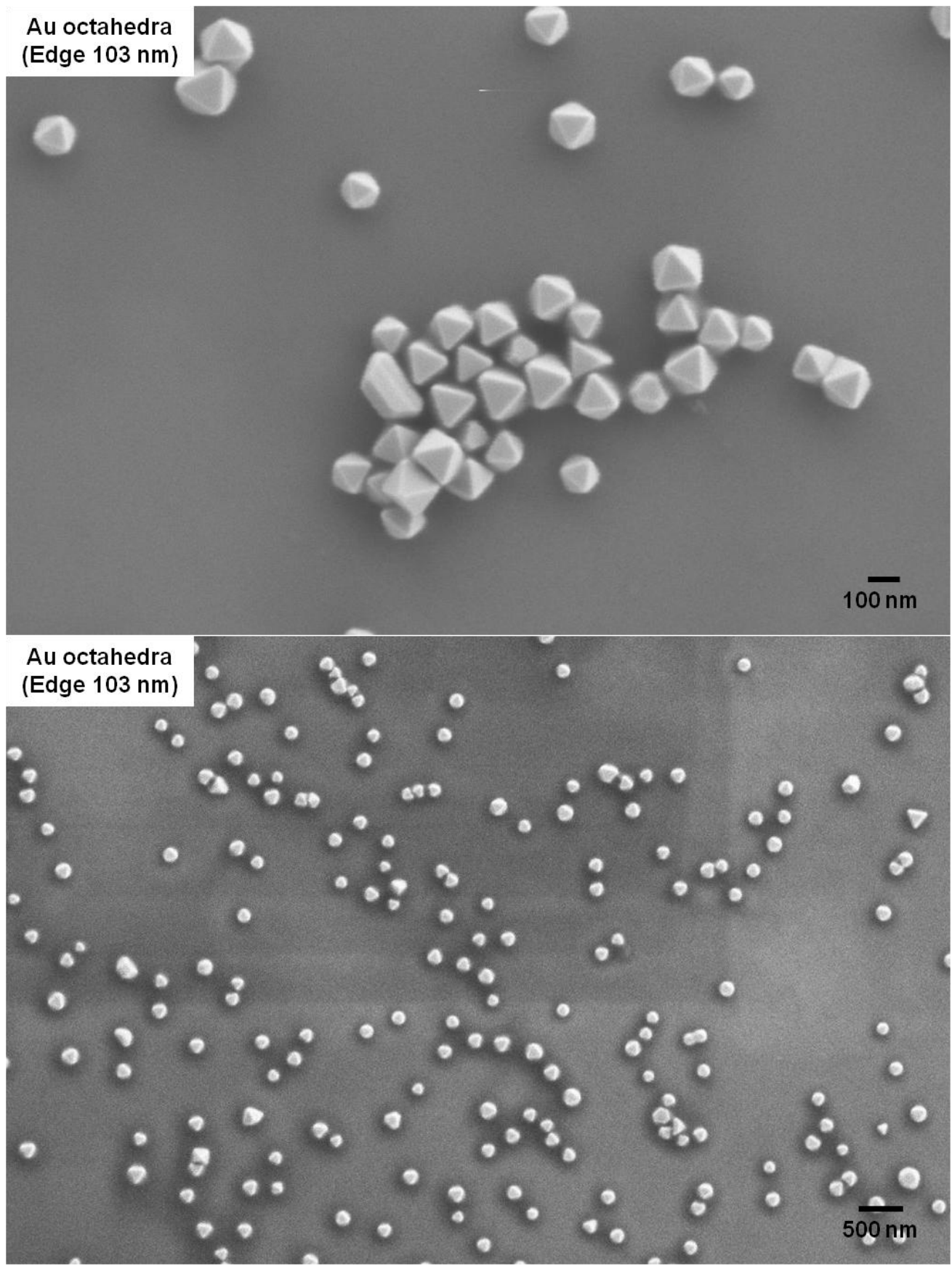

Figure S7. (b) SEM images of Au octahedra. Average edge length: 103 nm ± 18 nm.

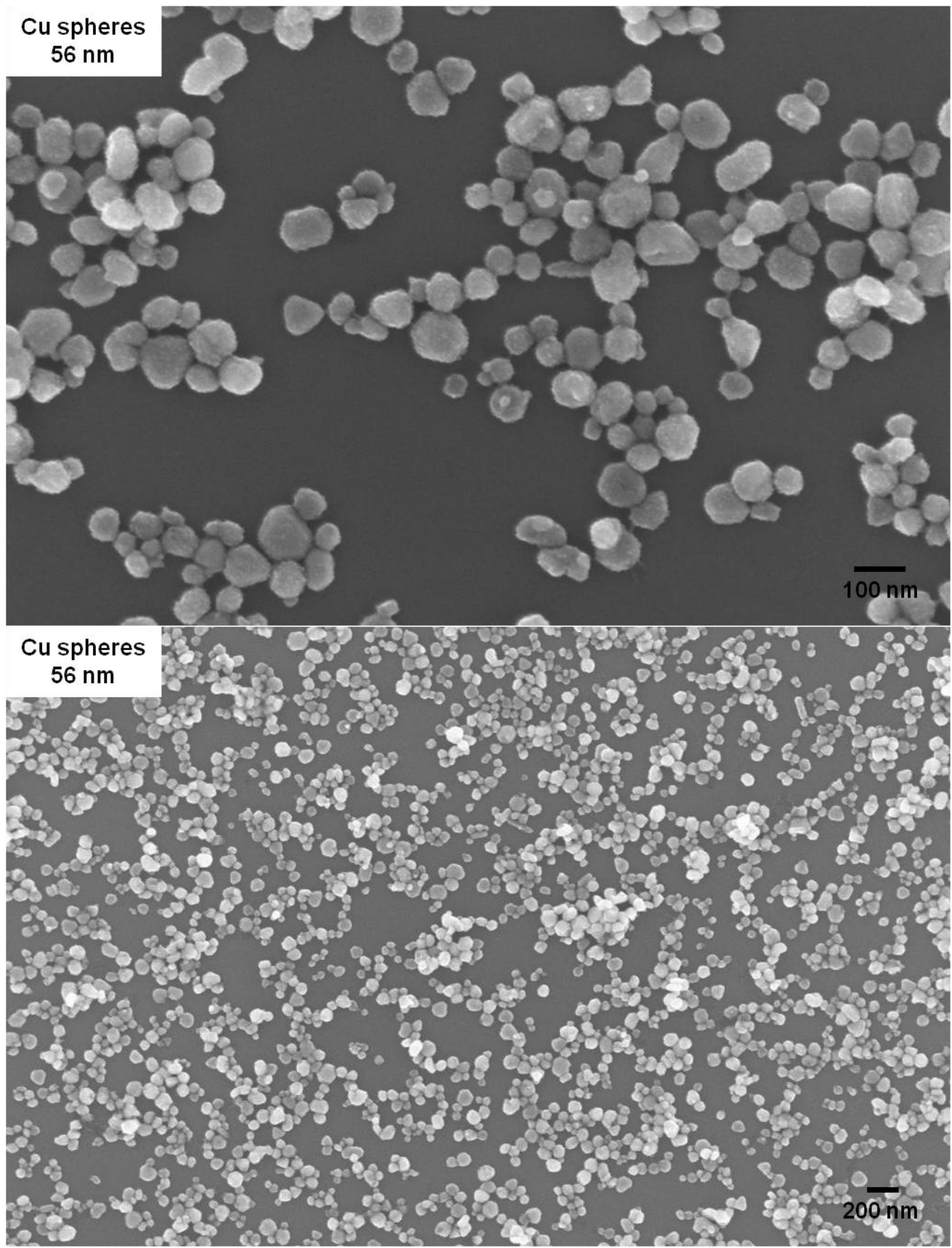

Figure S8. SEM images of Cu spheres. Average diameter by SEM analysis: 56 nm ± 18 nm. Size distribution by DLS indicates 84% of 94 nm ± 36 nm and 16% of 8 nm ± 2 nm.

S15

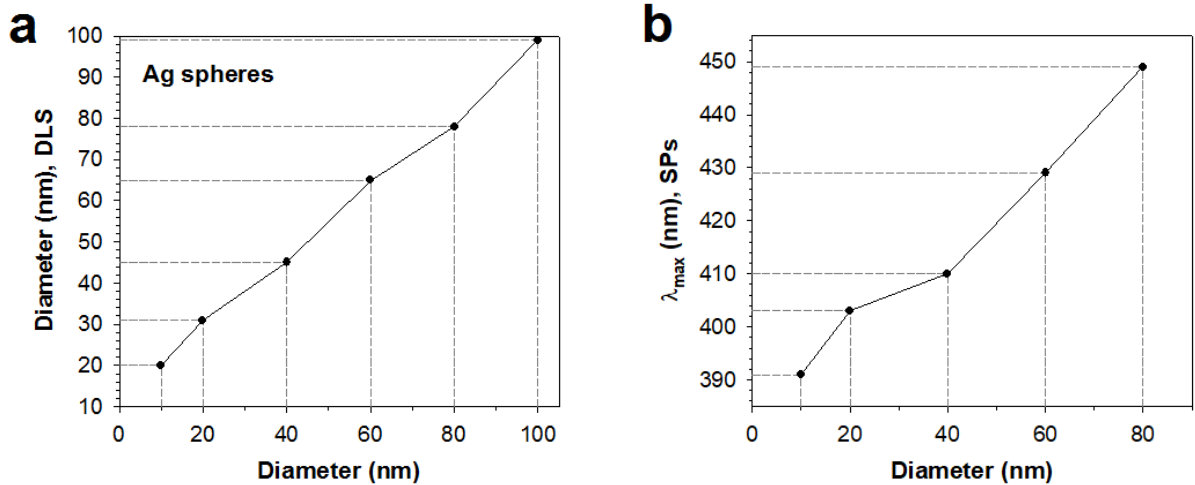

Figure S9. Average diameters of Ag spheres measured by (a) DLS z-average and (b) size-dependent $\lambda_{max}$ values of SPs.

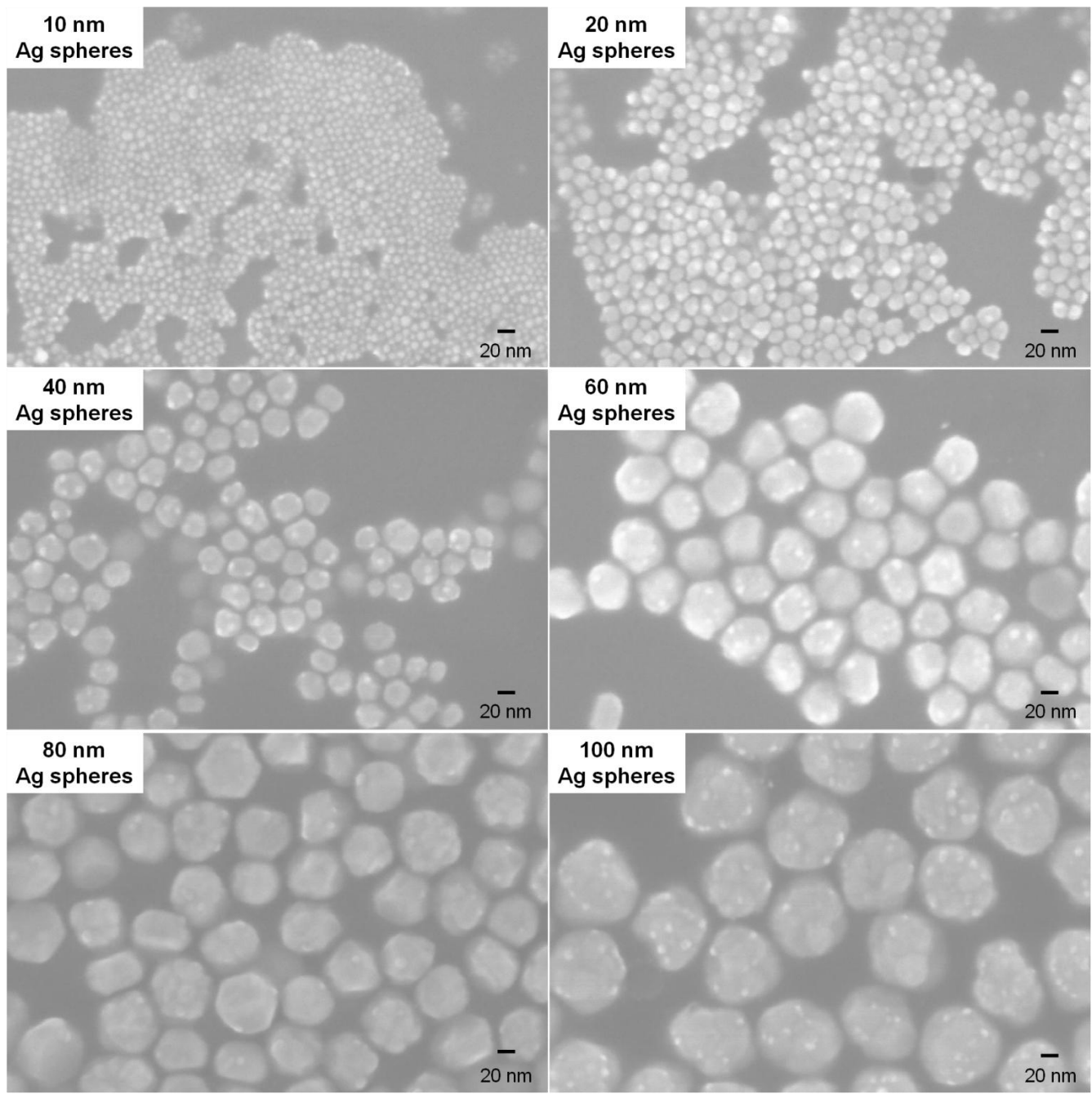

Figure S10. SEM image of Ag spheres (10, 20, 40, 60, 80, and 100 nm in diameter).

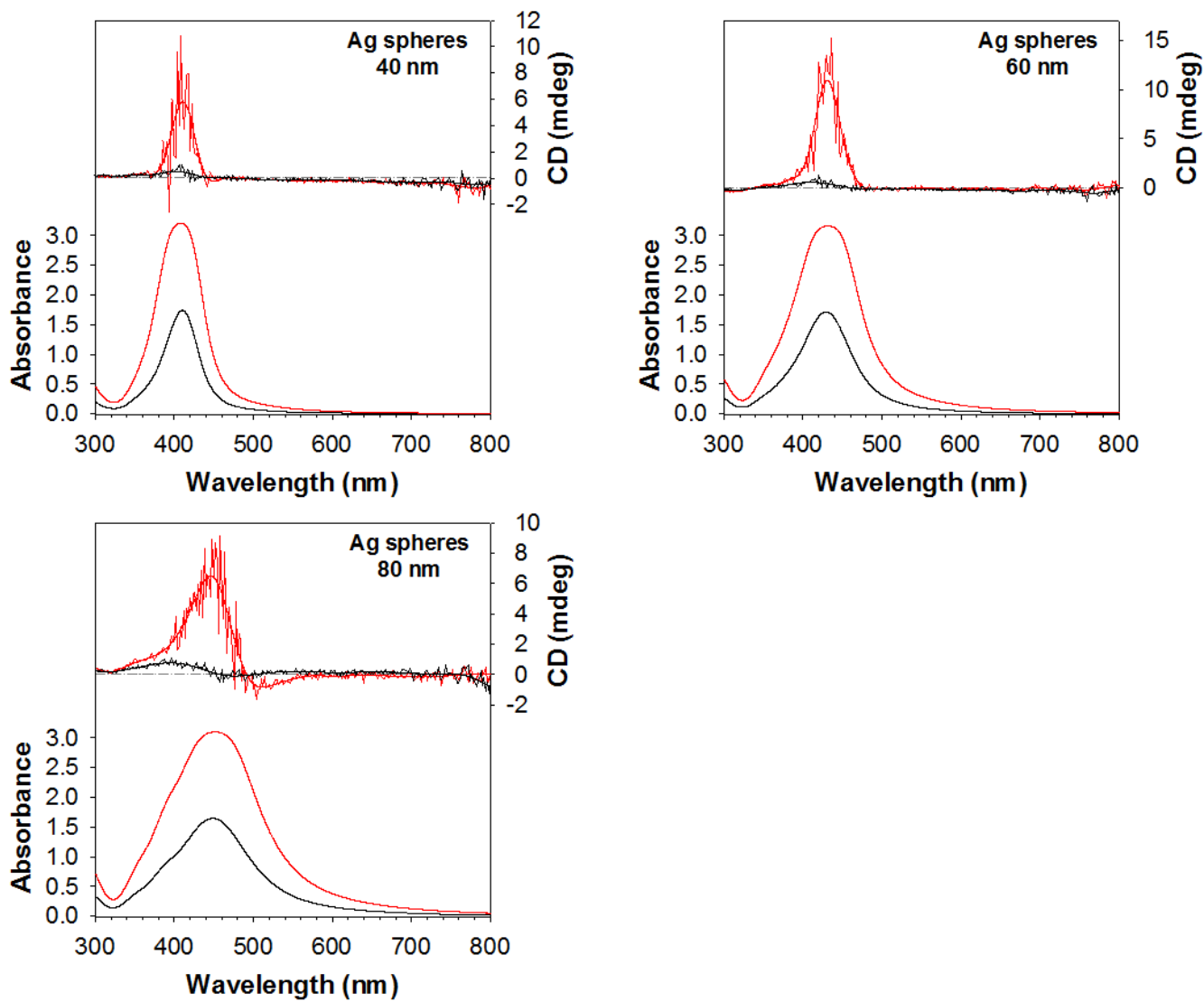

Figure S11. CD and UV-vis extinction spectra of Ag spheres (40, 60, and 80 nm in diameter). Path length: 1 cm.

**Determination of molar ellipticity for 100-nm silver spheres:**

Absorbance of 1.6 at the wavelength of 490 nm results in molarity = $6.4 \times 10^9$ particles/mL equal to $1.0 \times 10^{-11}$ moles/L = $1.0 \times 10^{-11}$ M ($3.7 \times 10^9$ particles/mL at Max OD = 0.93, provided by Ted Pella, Inc.; 99.4 nm ± 7.0 nm) ellipticity: (+) 0.9 mdeg at 405 nm and (-) 1.3 mdeg at 490 nm, and path-length is $10^{-2}$ m.

Thus, molar ellipticity is determined as in the following calculations,

at 405 nm, (+) 0.9 mdeg/($1.0 \times 10^{-11}$ M × $10^{-2}$ m) = (+) $9.0 \times 10^{12}$ mdeg·M$^{-1}$·m$^{-1}$

at 490 nm, (-) 1.3 mdeg/($1.0 \times 10^{-11}$ M × $10^{-2}$ m) = (-) $1.3 \times 10^{13}$ mdeg·M$^{-1}$·m$^{-1}$

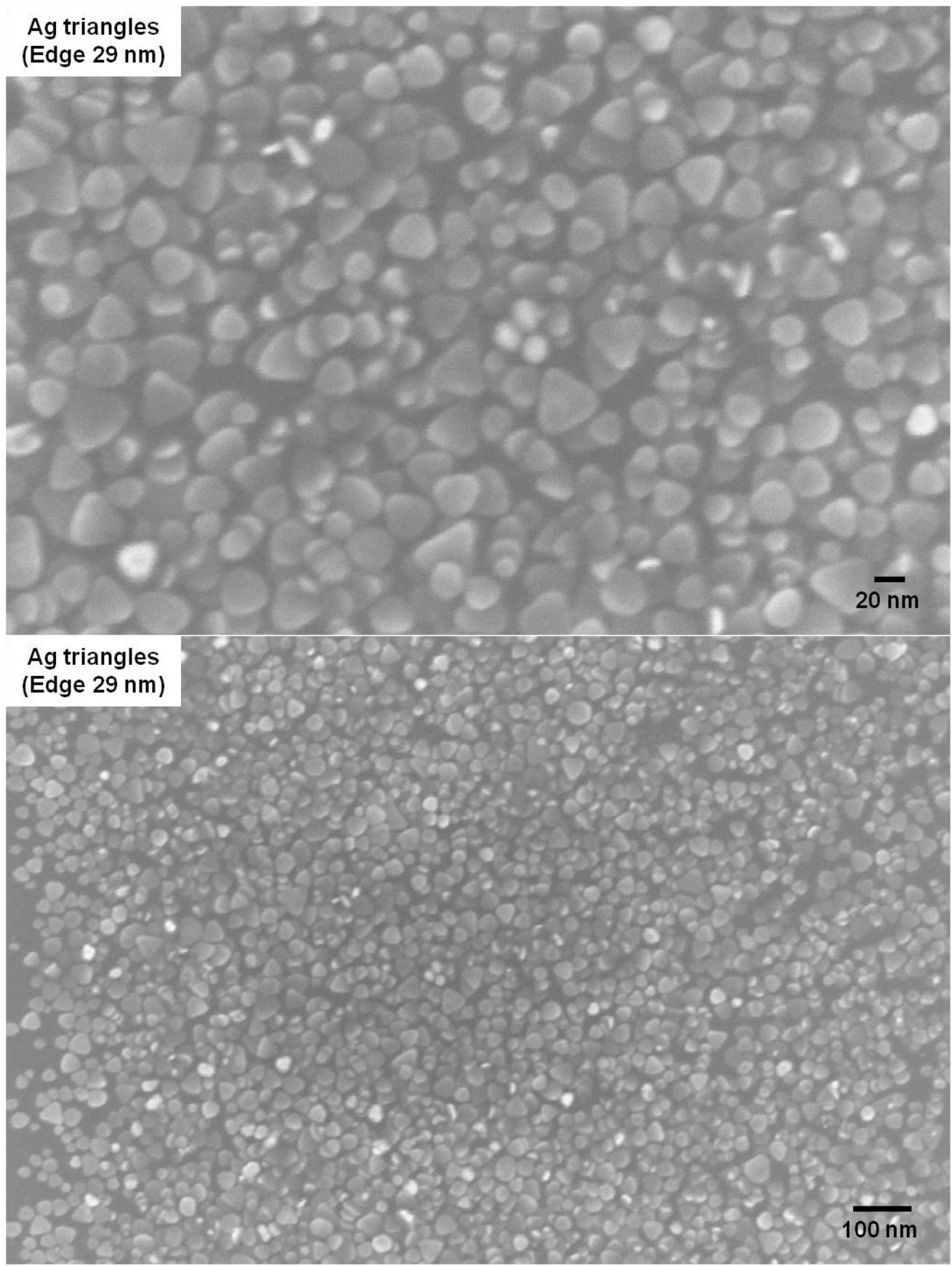

Figure S12. (a) SEM image of Ag triangles. Average edge length: 29 nm ± 6 nm; thickness < 10 nm.

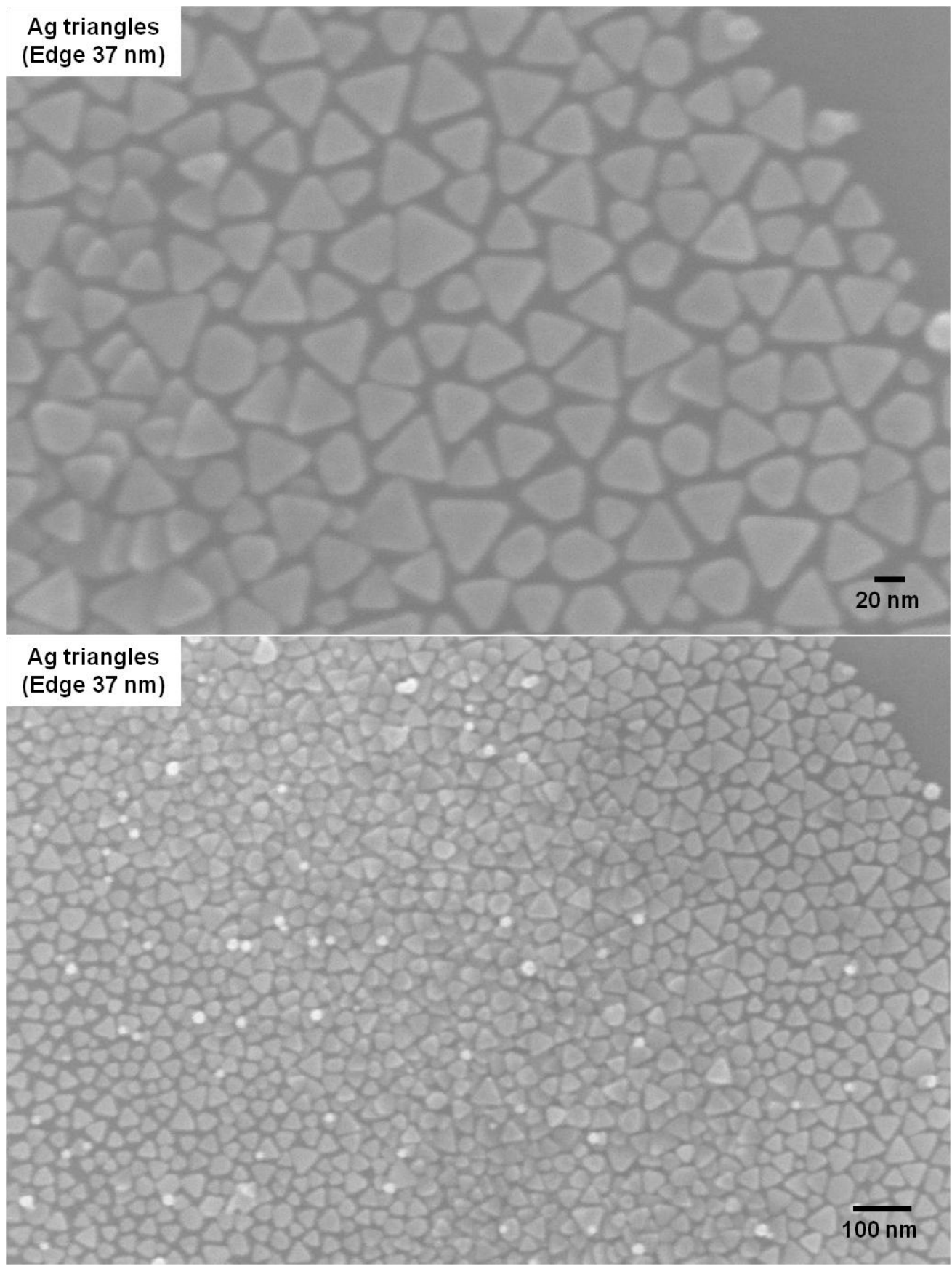

Figure S12. (b) SEM image of Ag triangles. Average edge length: 37 nm ± 7 nm; thickness < 10 nm.

**Discussion S3. electron spin-polarization:** Electron spin polarization may induce optical activity.[22] Spin polarization on noble metal surfaces occurs in surface states, and extent of spin-polarization depends on surface crystallographic structure. Shockley surface states are electronic states spatially located at outermost several metallic layers, and their energy regions are between the d-band and the Fermi level. Spin-orbit splitting in the surface state lifts spin degeneracy,[23] often called the Rashba-Bychkov effect.[24,25,26,27] Anisotropic surface states are well-known for gold, silver, and copper planar surfaces,[28,29,30] edge surfaces (Au,[31] Ag[32]) and nanoparticles (Au,[33] Ag[34]). For example, anisotropic dispersion of surface plasmons in single Ag crystals was attributed to the different effective mass of electrons of Shockley surface states on different crystal facets.[35] Electrons in the surface states exhibit a free-electron-like behavior, thus potentially they may couples to the conduction electrons of surface plasmons (SPs), where surface states are energetically overlapped with a bulk continuum.[30, 36] In particular, one may argue that upon plasmon excitation the quasiparticle interaction of electrons between the sp-conduction band and a spin-split surface state may be preferred, leading to anisotropic interfacial electron-electron scattering[37,38,39] and thus resulting in partially spin-polarized SPs, for to say < 1%. It has been recently reported that spin rotation with length of the order of tens of nanometers occurs because of multiple scattering of the surface state electrons on Au(111).[40] Alternatively, SPs may simply couple to the orbital angular momentum of the spin-split surface state, generating the total angular momentum (pseudospin).[41] While the spin-orbit coupling is relatively weak for those noble metals, the enhanced electric field by surface plasmon resonances may widen the energy gap substantially (Stark effects). According to the theory discussed above, CD response of Ag nanoparticles would be weaker than that of Au nanoparticles because of the negligible spin-orbit coupling for Ag,[28,29] which is opposed to the experimental observation of CD responses in this study. We found that molar ellipticities of spherical Ag and Au nanoparticles are comparable. Therefore, we conclude that a surface-state induced spin-polarization of surface plasmons does not occur in gold and silver nanoparticles.

**Discussion S4. CD response of spherical gold and silver nanoparticles stabilized by chiral molecules:** Figure S13 presents CD spectra of Au and Ag spheres synthesized using chiral molecules (size analyses by SEM in Figure S14 including DLS measurement results described in the figure caption). We found that molecular chirality is not related to the measured CD ellipticity of nanoparticles, consistent with other experimental results reported in the literature that chirality transfer from small molecules to a nanoparticle diminishes as the metal core size increases. When the core size exceeds ~2 nm in diameter, the amplitude of CD ellipticity is unnoticeable not only in the plasmon wavelength region,[42, 43] but also in the near-UV region.[43] In our study, CD ellipticity is dependent only on nanoparticle size regardless of which enantiomer is adsorbed on the surface of nanoparticles. First, Au spheres with a diameter of ~60 nm, which are stabilized by D-tryptophan (Trp) or L-Trp, exhibit only positive ellipticities near the plasmon wavelength centered at 580 nm, because of their large sizes acquiring quadrupolar SPs that have the positive rotational strength (Figure S13a). In the UV region, however, each CD spectrum of Trp-Au spheres displays an enantiomer-dependent amplitude of ellipticity. Taking the slope of the positive ellipticity of interband transitions at the wavelength of 250 – 500 nm region as a baseline, a mirror image of ellipticity becomes obvious particularly in the 250 – 350 nm wavelength range with an ascending peak of L-Trp Au spheres and a descending valley of D-Trp Au spheres. This mirror-image indicates that chirality of Trp adsorbed on the nanoparticle surface does transfer to d-electrons[44] of the metal rather than conduction electrons, reflecting light absorption of Trp in the UV region. Figure S13b shows CD and UV-vis extinction spectra of Au spheres synthesized using L-ascorbic acids, and Figure S13c shows CD and UV-vis extinction spectra of Au spheres synthesized using tannic acids. Both molecules are commonly used in preparing Au spheres. For these small Au spheres with an average diameter < 30 nm, only negative ellipticities are obtained at the plasmon wavelength of around 520 nm, because of their small sizes fitting only dipolar SPs that have the negative rotational strength, similar to the citrate-stabilized Au spheres having the similar sizes. In the UV regions, ellipticity sign of the Au spheres corresponds to the molecular chirality of L-ascrobic acids and tannic acids, respectively, indicated by the ellipticity sign of L-ascrobic acid or tannic acid solutions without nanoparticles. Moreover, the ellipticity amplitude of Ag spheres stabilized by the chiral tannic acids is similar to that stabilized by the achrial citrate anions (Figure S13d), also indicating no relation between molecular chirality and the observed CD at the plasmon wavelength. These experimental results confirm the observed CD of SPs is related to only the size of nanoparticles, although these results are in conflict with to the reported examples of plasmon-enhanced CD for individual Ag and Au nanoparticles, not for assembled nanoparticles (see **Discussion S5,** below).

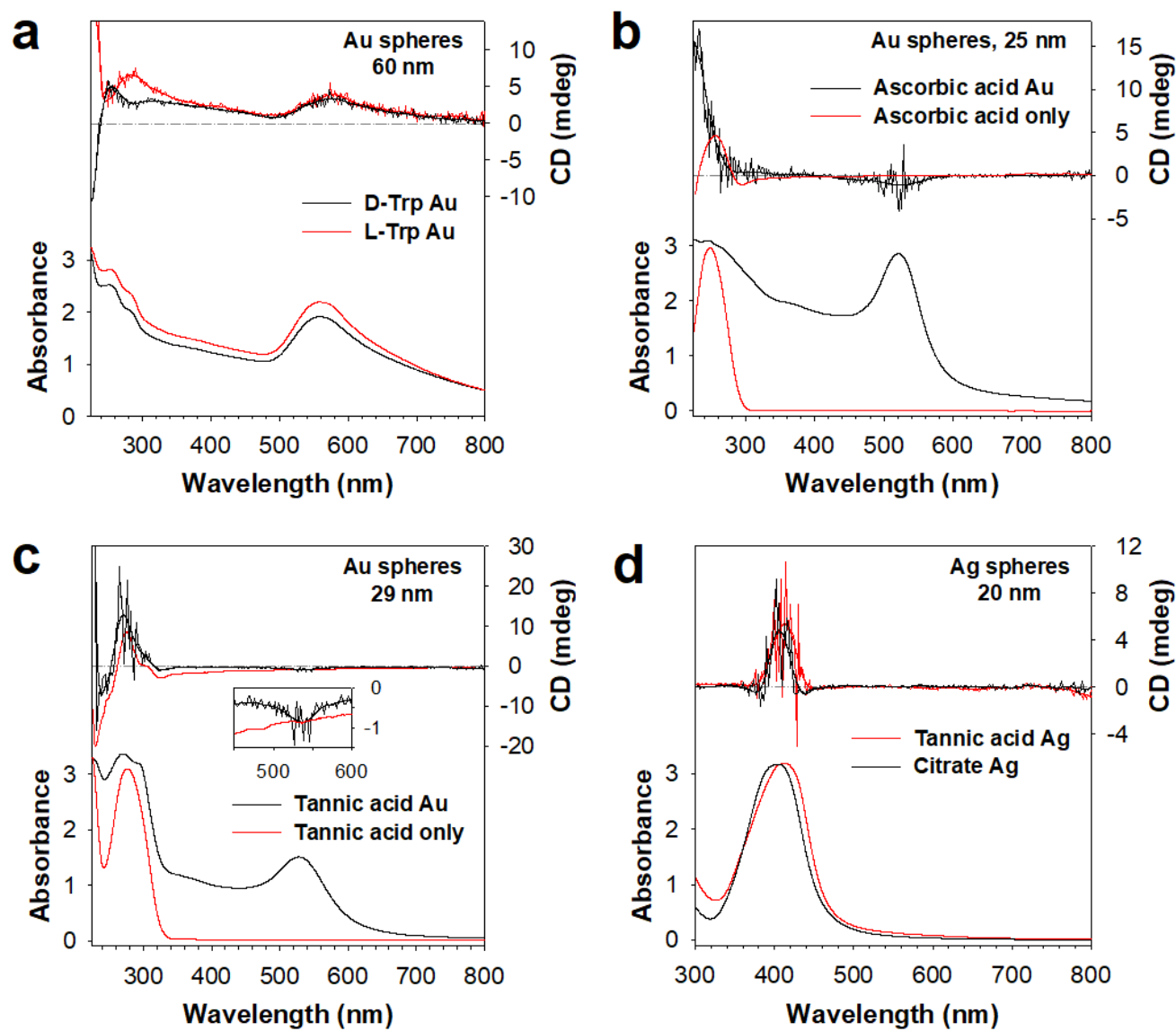

Figure S13. CD and UV-vis extinction spectra of Au and Ag nanoparticles stabilized by chiral surfactants. (a) D(L)-tryptophan (Trp) Au spheres, (b) Ascorbic acid Au spheres, ascorbic acid only: 0.4 mM, (c) Tannic acid Au spheres, tannic acid only: 0.02 mM, (d) Tannic acid Ag spheres compared with citrate Ag spheres. Path length: 1 cm.

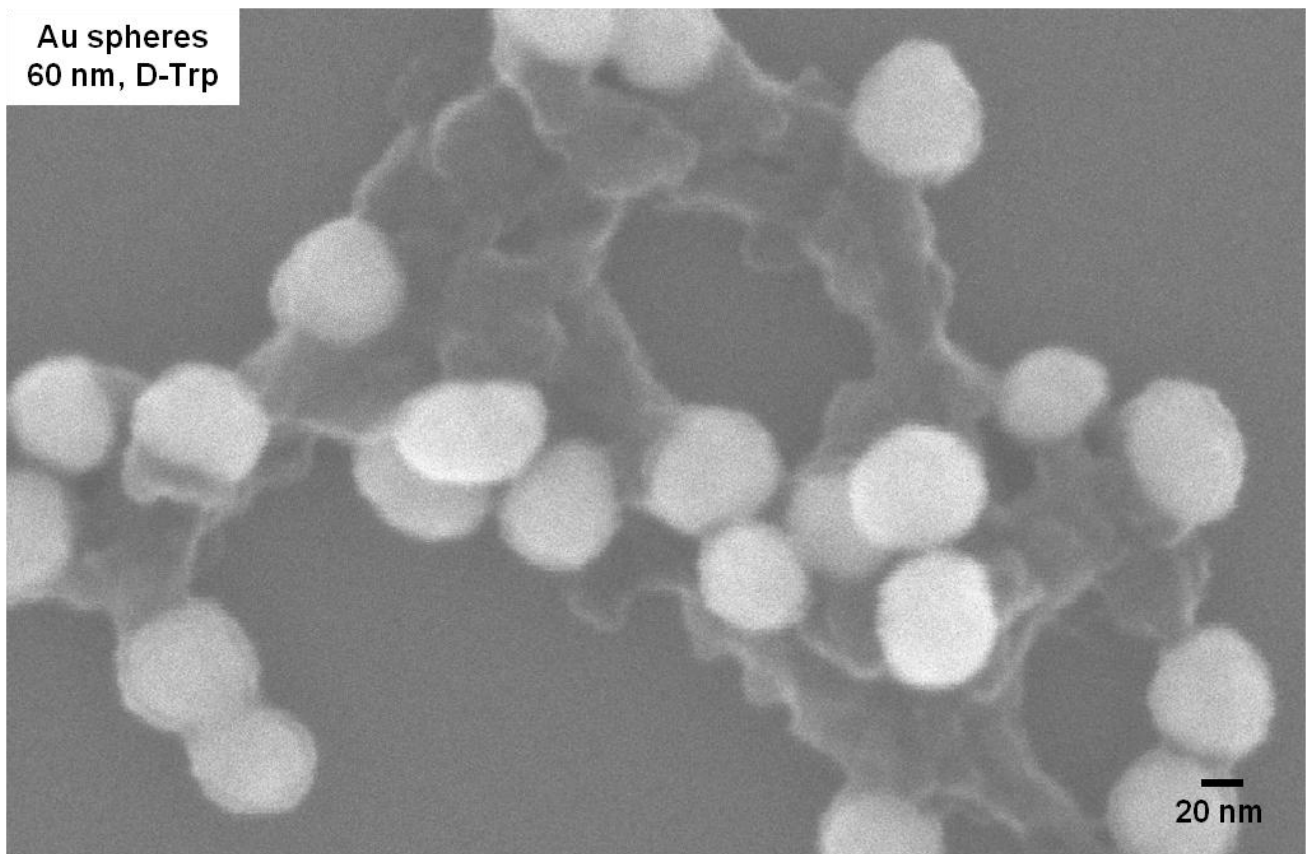
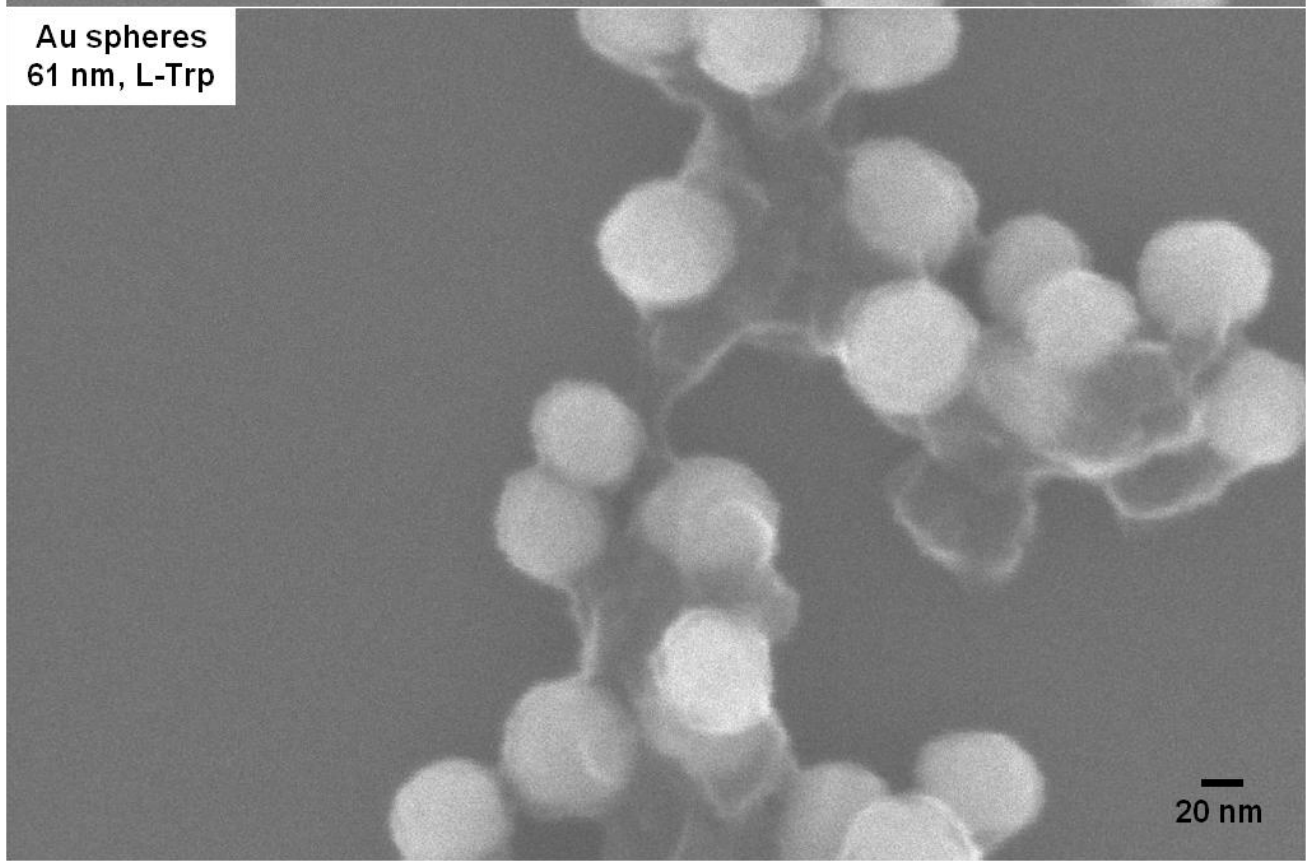

Figure S14. (a) SEM images of Au spheres synthesized using D/L-tryptophan (Trp). Average diameter: D-Trp-Au 60 nm ± 7 nm, L-Trp-Au 61 nm ± 7 nm.

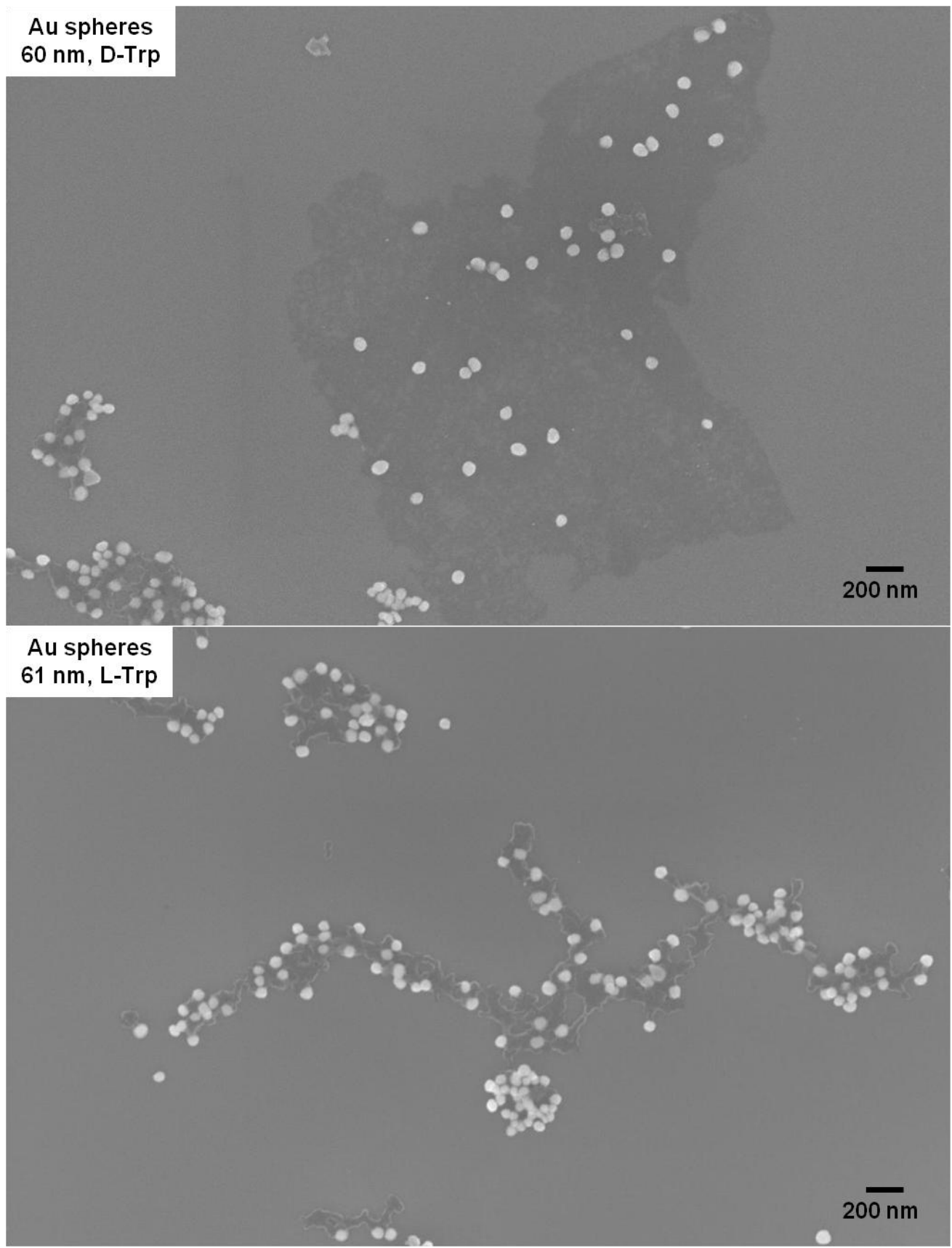

Figure S14. (a) (continued) SEM images of Au spheres synthesized using D/L-tryptophan (Trp).

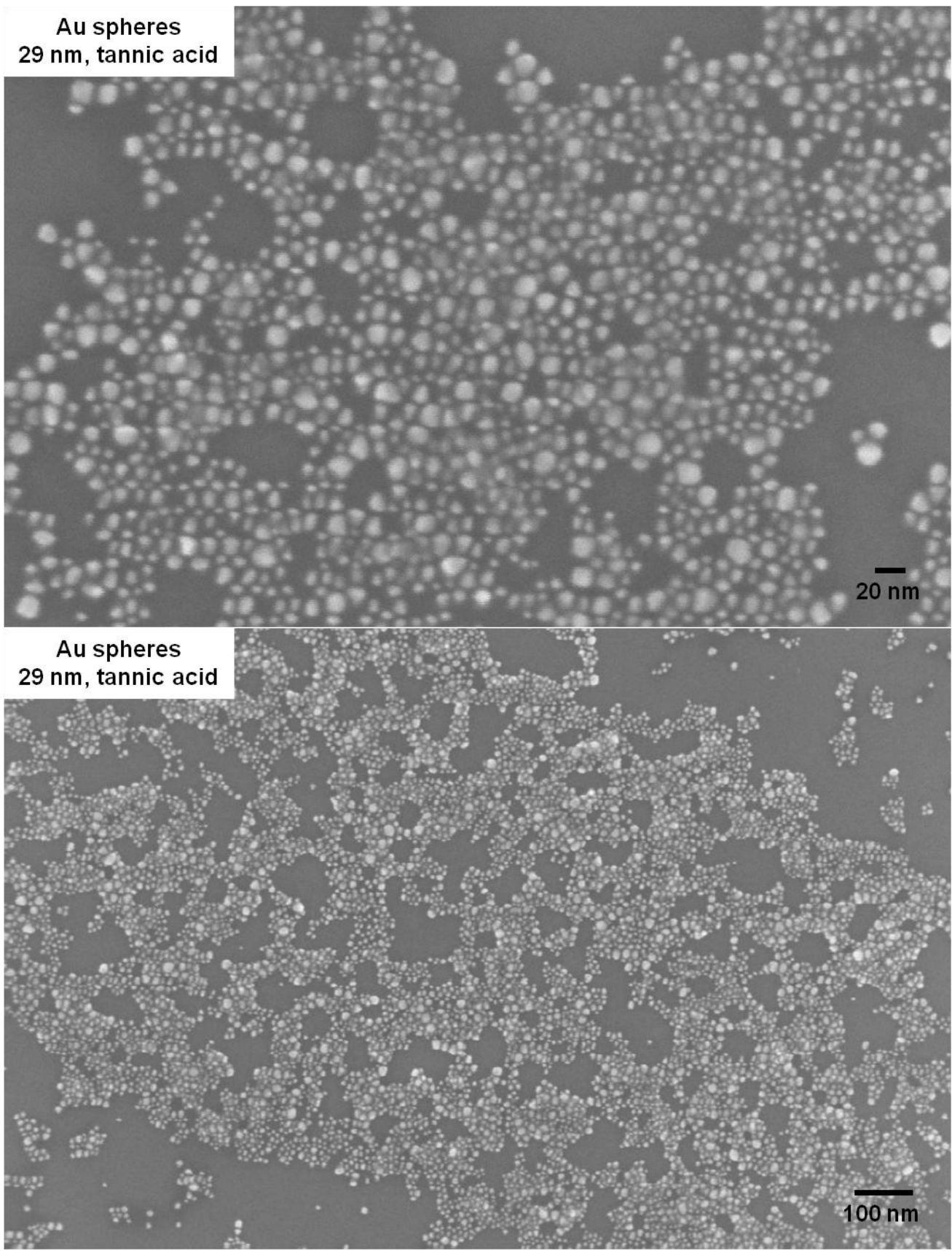

Figure S14. (b) SEM images of Au spheres synthesized using tannic acid. Average diameter measured by DLS: 29 nm.

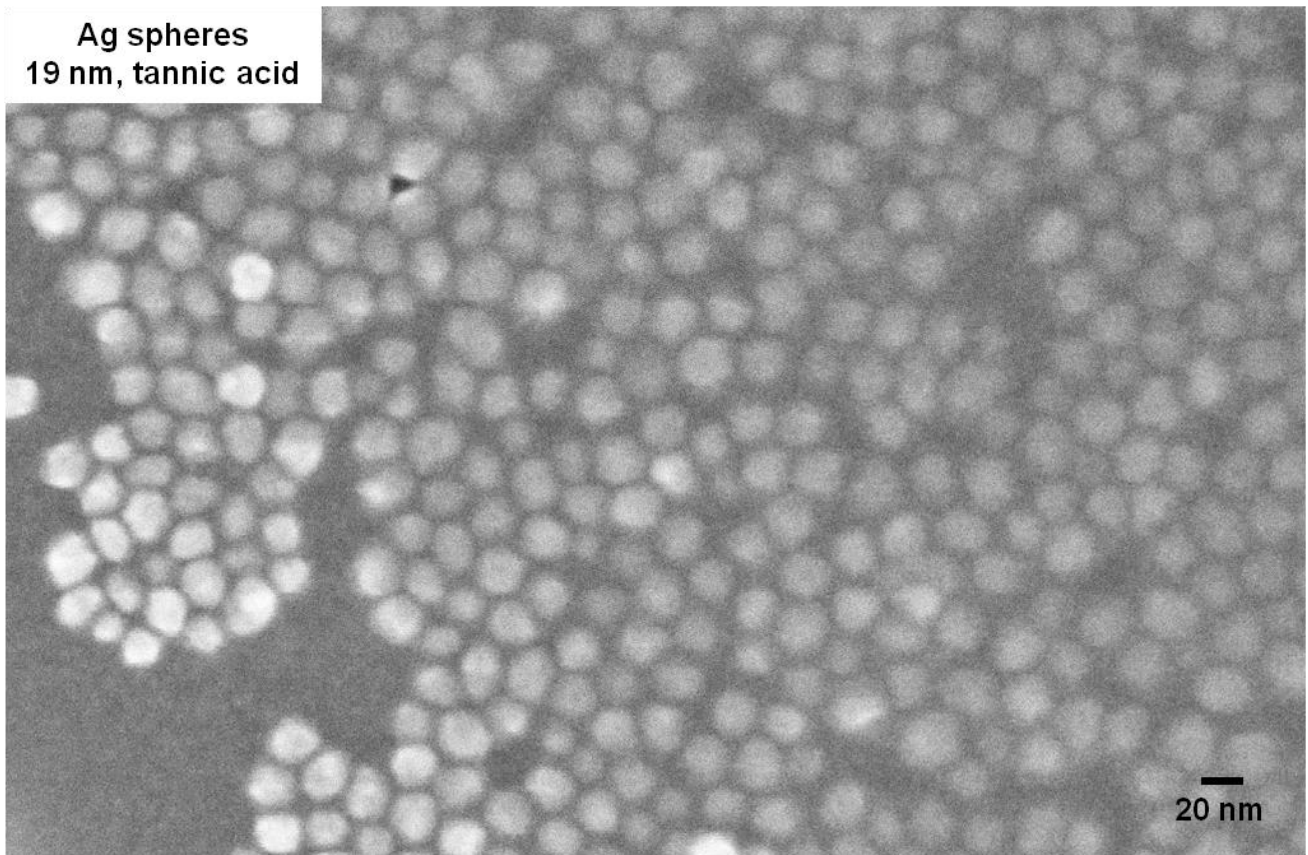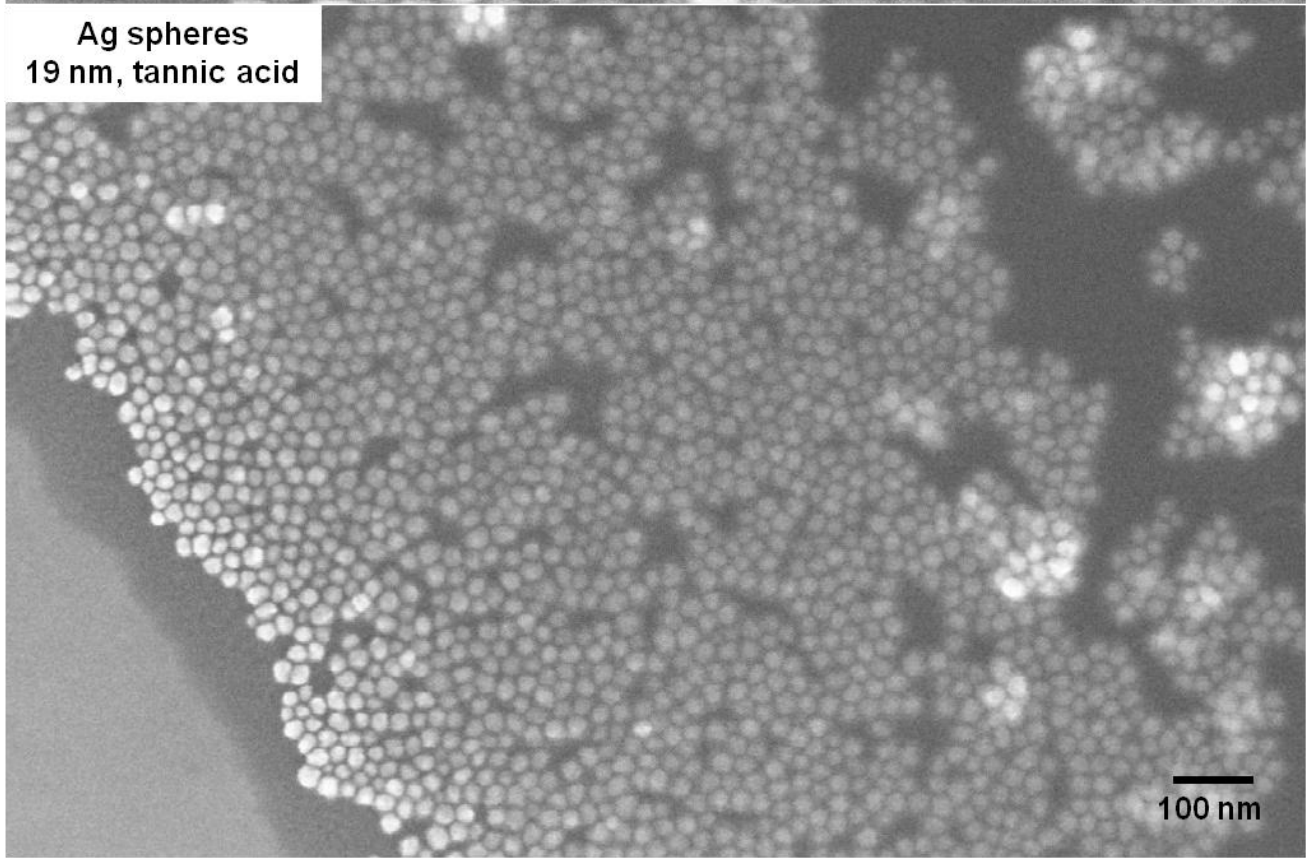

Figure S14. (c) SEM images of Ag spheres stabilized by tannic acid. Average diameter: 19 nm ± 3 nm.

**Discussion S5. Plasmon-enhancement of chirality from macromolecules absorbed on nanoparticles:** Chiral transfer from ligand to metal nanoparticles has gained great attention for a past decade. Since the observation of circular dichroism of ligand-protected gold nanoclusters was made by Schaaff and Whetten in 2000,[45] optical activity for metal nanoclusters with a diameter less than 2 nm has been well-understood by both experiments and computation.[46,47,48] The CD response of the metal nanoclusters is a consequence of metal-ligand orbital hybridization and interplay between cluster-core distortion and ligand dissymmetry.[49,50] For plasmonic gold and silver nanoparticles with a diameter larger than 2 nm,[51,52] however, dissymmetry of adsorbed chiral molecules is no longer transferred to the whole body of the metallic core, which is consistent with the experimental observation of the absence of CD bands in the surface plasmon energy region.[43,53,54] In contrast, a recent theory indicates that plasmon-induced chirality may occur in large metal nanoparticles via Coulomb interactions between chiral ligands and surface plasmons.[55] This CD mechanism has been demonstrated for Ag[56,57,58] and Au[59] nanoparticles in part, only when nanoparticles are functionalized with macromolecules including DNA, peptides, and riboflavin. Until now, an experimental result of the plasmon-enhanced CD of individual nanoparticles functionalized with small chiral molecules such as amino acids or hydroxy acids (e.g., lactic acid, glyceric acid, and tartaric acid; these molecules with carboxylic groups deprotonated will not cause nanoparticle aggregation upon addition to aqueous solution of the commonly used citrate-stabilized Au or Ag nanoparticles) has never been reported, although chiral molecules can induce chiral assembly of individual nanoparticles and thus produce strong scattering CD signal in the far-field CD spectroscopy. The CD theory of plasmon-enhancement also predicts that a uniform distribution of chiral molecules on a nanoparticle surface will produce a net zero CD response, because of the fact that polarity sign at the electron-cloud side is opposite to that at the jellium-background side in the given direction of the electric field of dipolar SPs,[21] cancelling out the local interface CD responses.[60,61] In other words, the plasmon-enhanced CD mechanism is origin-dependent and the net CD signal must vanish in solution. Nonetheless, the large magnitude of polarizability of the macromolecules mentioned above (i.e., DNA, peptide, riboflavin) may play a crucial role in symmetry-breaking at individual nanoparticles upon functionalization, probably through inhomogeneous distribution of molecules on the nanoparticle surface.[62,63] In addition, strength of the Coulomb interaction between molecules and surface plasmons is unknown, and electron exchange/correlation, which are negligible in the one-electron model of molecular chirality, should be considered in the many-body system of electrons in a metal nanoparticle.[37,64] It is surprising that studies on the optical absorption of individual molecules (not of J-aggregates[65,66]) on plasmonic nanoparticle surfaces are rare,[67] suggesting that Coulomb interactions between the conduction electrons in metal and bound electrons in molecules adsorbed on nanoparticles have not been well-understood so far. We emphasize that experimental results about plasmon-enhanced CD for individual nanoparticles have been limited for nanoparticles functionalized with relatively large molecules such as DNA.

**Discussion S6. Effects of surface charge and solvent:** Since the Au and Ag spheres used in this study are negatively-charged by citrate molecules and stabilized in water, Au spheres stabilized by positively-charged molecules, other anions and solvents have also been examined in order to study the effects of surface charge and solvent as well as refractive index. First, we synthesized Au spheres using acrylate and polyethylenimine (PEI). The former produces negatively-charged nanoparticles whereas the letter produces positively-charged ones, confirmed by zeta-potential measurements (Table S3). Using those molecules usually produces small nanoparticles with narrow size distribution (Figure S15). Second, we synthesized PVP-stabilized Au spheres using $NaBH_4$ as a reducing agent, which the surface charge is expected to be nearly neutral because of the neutral charge of PVP, although remaining $BH_4^-$ ions led to a negative surface charge. Third, we functionalized citrate-stabilized Au spheres (average diameter by DLS: 26 nm), which were synthesized for the study of charge and solvent effects, with mercaptopropionic acid in water to examine the surface effect[68] of gold-thiolate bonds on optical activity. Last, we redispersed citrate-stabilized Au spheres, which were sedimented from the aqueous solution by centrifugation, in ethanol in order to examine any unusual effect of water confined on the nanoparticle surface[69] as well as the effect of refractive index. Figure S16 shows CD and UV-vis extinction spectra of those Au spheres with diameters < 50 nm. The CD spectra indicate CD response of those Au spheres is dependent on nanoparticle size and all of the small nanoparticles exhibit only negative ellipticities of dipolar SPs, irrelevant to surface charge and solvent. Interestingly, the PEI-stabilized Au spheres display a unusual band shape with one peak and two valleys around the plasmon wavelength centered at 530 nm. This may be an indication of the resonance condition in the longitudinal-transverse plasmon coupling through change of refractive index, probably resulting in a destructive Fano resonance (i.e., the incident wave is not distorted while a constructive Fano-like resonance results in an elliptical light-polarization).[70]

**Table S3. Zeta-potential values of Au spheres functionalized with carboxylate, ammonium, or ketone anchor groups**

| Reducing agent | Functional group | Zeta potential (mV) |
| --- | --- | --- |
| Citrate | Carboxylate (–COO$^-$) | -40.6 ± 23.0 |
| Acrylate | Carboxylate (–COO$^-$) | -33.4 ± 27.0 |
| Polyethylenimine (PEI) | Ammonium (–NH$_3^+$) | +17.1 ± 11.3 |
| Polyvinylpyrrolidone (PVP) + BH$_4^-$ | Cycloketone (–C=O) | -19.6 ± 16.4 |

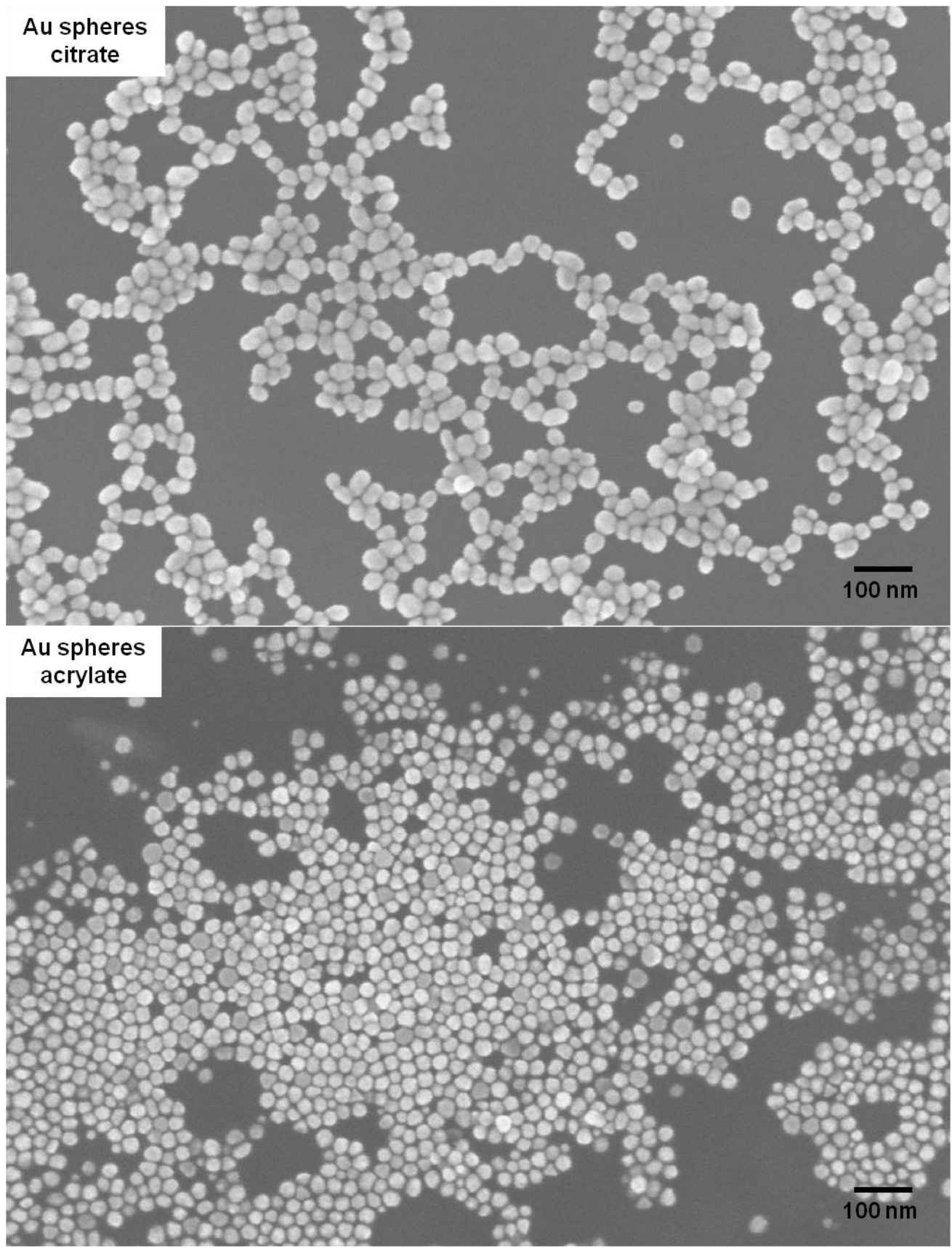

Figure S15. (a) SEM images of Au spheres synthesized using citrate and acrylate. Average diameters measured by DLS: citrate-Au 26 nm and acrylate-Au 36 nm.

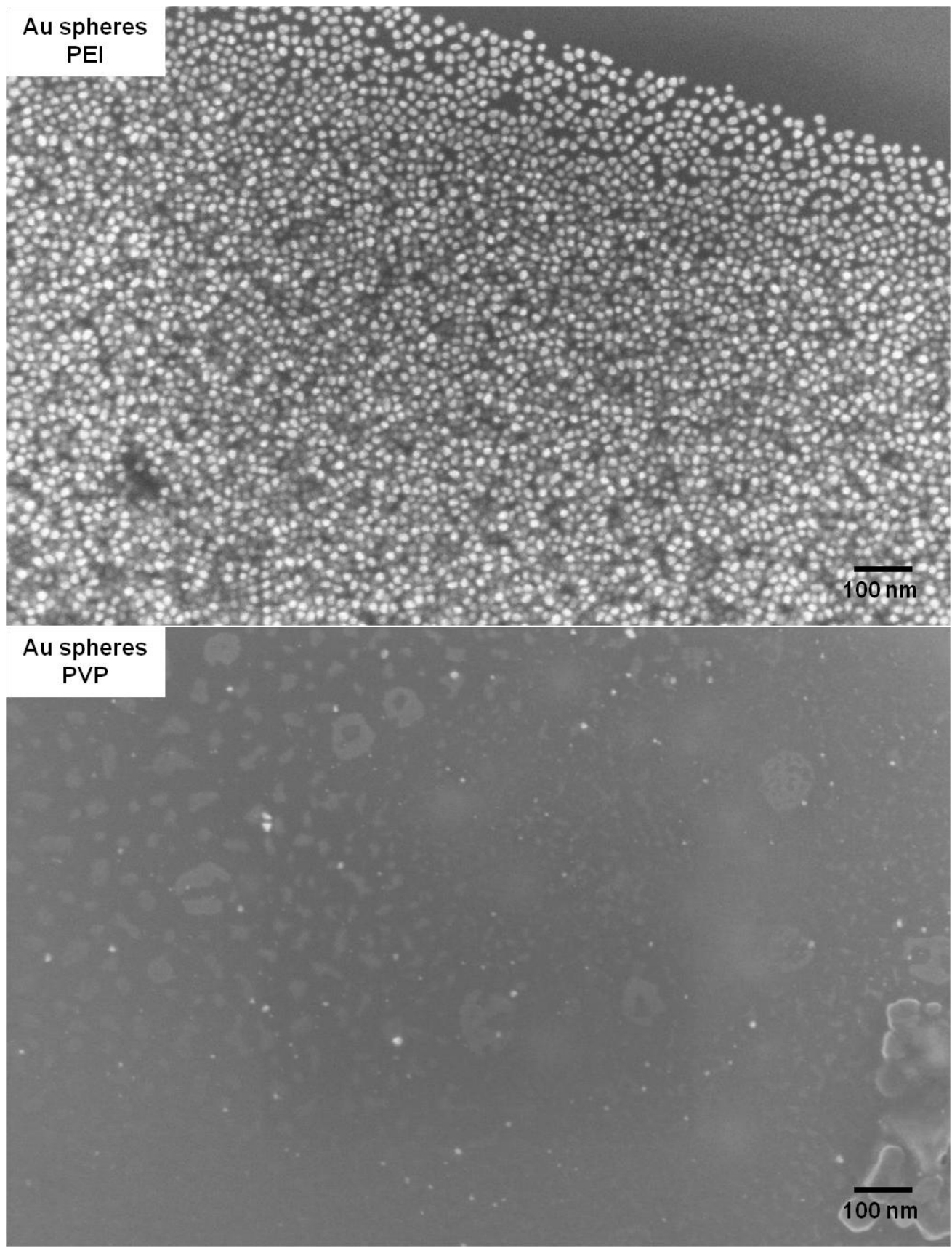

Figure S15. (b) SEM images of Au spheres synthesized using polethyleneimine (PEI) and polyvinylpyrrolidone (PVP)/$NaBH_4$. Average diameters measured by DLS: PEI-Au 38 nm and PVP-Au 12 nm.

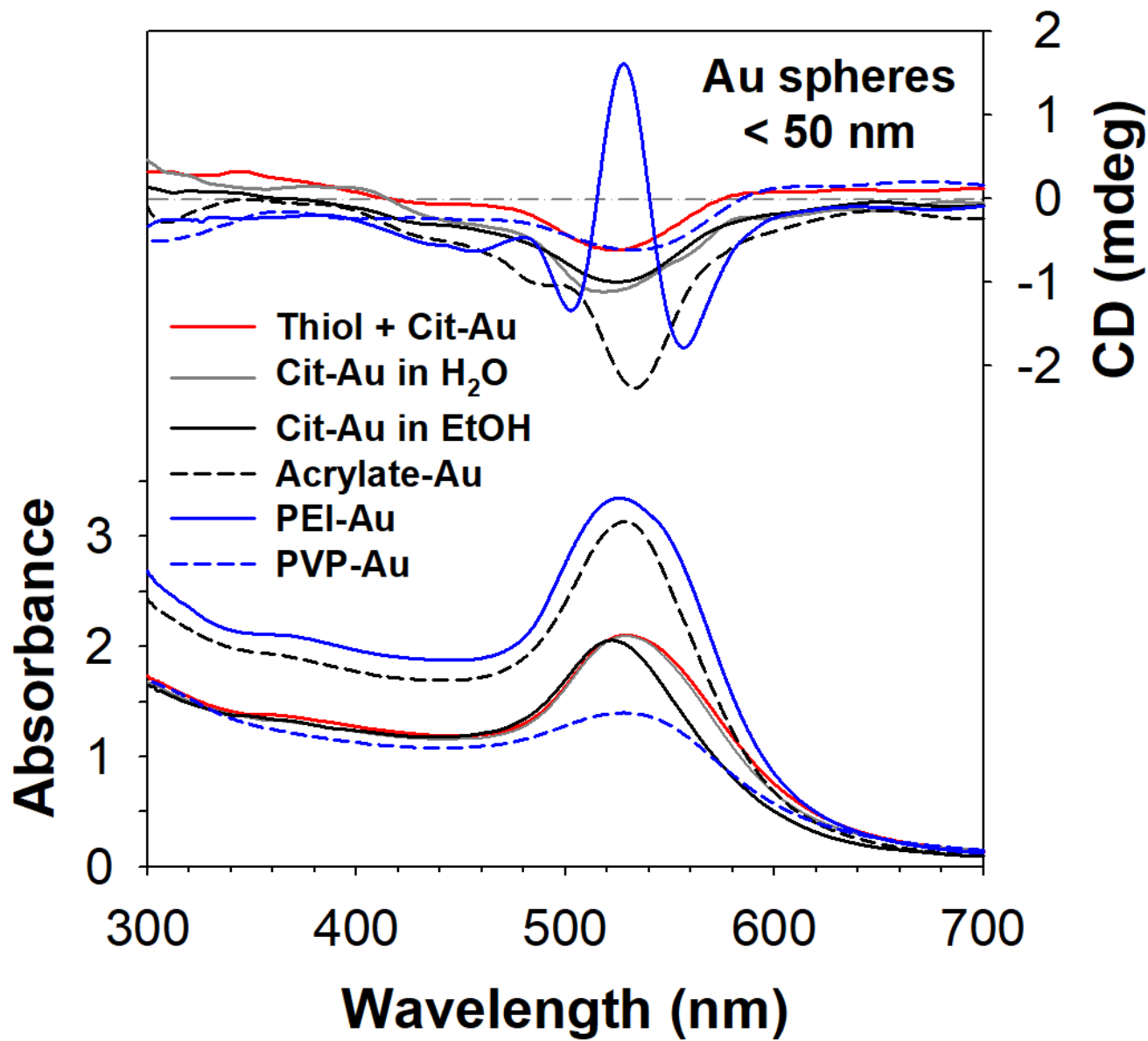

Figure S16. CD and UV-vis extinction spectra of spherical Au nanoparticles stabilized by achiral molecules (mercaptopropionic acid, citrate, acrylate, PEI, PVP) and stabilized in a different solvent (ethanol). Average diameters measured by DLS: citrate-Au 26 nm, acrylate-Au 36 nm, PEI-Au 38 nm, and PVP-Au 12 nm. Path length: 1 cm.

**Supplemental References**